\documentclass[prd,twocolumn,superscriptaddress,showpacs,groupedaddress]{revtex4-2}
\usepackage{hyperref}
\usepackage{physics}
\usepackage{dsfont}
\usepackage{amsfonts}
\usepackage{bm}
\usepackage{graphicx}
\usepackage{xcolor}
\usepackage{mathtools}
\usepackage{verbatim}
\usepackage{tikz}
\usepackage[shortlabels]{enumitem}

\usepackage{dsfont}
\usepackage{bbold}

\begin{document}

\title{Negative tripartite mutual information after quantum quenches in integrable systems}
\date{\today}
\author{Fabio Caceffo}
\affiliation{Dipartimento di Fisica dell'Universit\`a di Pisa and INFN, Sezione di Pisa, I-56127 Pisa, Italy}
\author{Vincenzo Alba}
\affiliation{Dipartimento di Fisica dell'Universit\`a di Pisa and INFN, Sezione di Pisa, I-56127 Pisa, Italy}

\begin{abstract}
We build the  quasiparticle picture for the tripartite mutual information (TMI) after quantum 
quenches in spin chains that can be mapped onto free-fermion theories. A nonzero TMI (equivalently, 
topological entropy) signals quantum correlations between three regions of a quantum many-body system. 
The TMI is sensitive to entangled multiplets of more than two quasiparticles, i.e., beyond the 
entangled-pair paradigm of the standard quasiparticle picture. 
Surprisingly, for some nontrivially entangled multiplets the TMI is negative at intermediate times. 
This means that the mutual information is monogamous, similar to holographic theories. 
Oppositely, for multiplets that are ``classically'' entangled, the TMI is positive. 
Crucially, a negative TMI reflects that  the entanglement content of the multiplets is 
not directly related to the Generalized Gibbs Ensemble (GGE) that describes the post-quench steady state. 
Thus, the TMI is the ideal lens to observe the weakening of the relationship between 
entanglement and thermodynamics. We benchmark our results in  the $XX$ chain and in the 
transverse field Ising chain. In the hydrodynamic limit of long times and large intervals, with their ratio 
fixed, exact lattice results are in agreement with the quasiparticle picture. 
\end{abstract}

\maketitle

%####################################################
\section{Introduction}
\label{sec:intro}

Recent years witnessed the stunning success of hydrodynamic approaches to 
describe entanglement dynamics in integrable quantum 
many-body systems. The so-called quasiparticle picture, 
which was originally put forward~\cite{calabrese2005evolution} in the context of Conformal Field Theory (CFT), 
spurred a tremendous amount of activity~\cite{fagotti2008evolution,alba2017entanglement,bertini2022growth,alba2021generalized}. 
The tenet of the quasiparticle picture is that in integrable systems 
after a quantum quench~\cite{calabrese2016introduction,vidmar2016generalized,essler2016quench} 
the entanglement growth is attributable to the ballistic propagation of entangled \emph{pairs} of quasiparticles. 
The quasiparticle picture proved to be succesful in generic quenches in free 
theories~\cite{fagotti2008evolution}, as well as 
in interacting integrable systems~\cite{alba2017entanglement,alba2018entanglement}. 
Crucially, the quasiparticle picture relies on 
thermodynamic information. Precisely, in the presence of entangled pairs, the quasiparticles 
and the entanglement shared between them 
are extracted from the Generalized Gibbs Ensemble~\cite{calabrese2016introduction} (GGE)
that describes the steady state after the quench. The only 
ingredient of non-thermodynamic origin is the pair structure itself, or, in general, the number of 
entangled quasiparticles generated after the quench. This entanglement pattern, i.e., the type of 
entangled multiparticle excitations that are resposible for the entanglement spreading, is enforced by 
the initial state. Interestingly, the pair structure is related  to 
a special class of ``integrable'' quenches~\cite{piroli2017what}, for which 
the GGE can be obtained  in closed form.

Here we study quantum quenches that give rise to entangled \emph{multiplets} of excitations, 
through the lens of the tripartite mutual information ($TMI$). 
An interesting quench producing entangled multiplets was already explored 
in Ref.~\cite{bertini2018entanglement}. In that setup, however, 
the entanglement content of the multiplet is fully determined by the 
GGE, and it is traced back to 
a classical-in-nature constraint between the quasiparticles forming the multiplet. 
Precisely, the focus of Ref.~\cite{bertini2018entanglement} was on 
quenches in  the so-called $XX$ chain (see section~\ref{sec:models}). The prequench 
initial states were obtained by repeating a unit cell of $n$ sites containing a 
single fermion. During the post-quench dynamics entangled multiplets 
formed by  $n$ quasiparticles are generated. However, as anticipated, the link between 
entanglement and thermodynamics is only mildly broken because the 
entanglement between the quasiparticles forming the multiplet can be 
obtained from the $GGE$ describing the steady state after 
the quench. We anticipate that this is reflected in the $TMI$ being positive.  
Interestingly, it was 
shown in Ref.~\cite{bastianello2018spreading} that it is possible to engineer quantum quenches 
giving rise to genuinely \emph{quantum-correlated} multiplets. 
Similar to Ref.~\cite{bertini2018entanglement}, entangled 
multiplets of quasiparticles are produced after the quench. The quasiparticles propagate 
ballistically, implying that a hydrodynamic  description of entanglement spreading is 
possible. However,  the entanglement shared between the quasiparticles cannot be extracted 
from the GGE. Hence, the link between entanglement dynamics and thermodynamics is 
strongly broken. As we will show in section~\ref{sec:ising} this leads to negative $TMI$. 
Still, even in the presence of multiplets (cf.~Ref.~\cite{bertini2018entanglement} and Ref.~\cite{bastianello2018spreading}) 
standard measures of bipartite 
entanglement~\cite{amico2008entanglement,calabrese2009entanglement,laflorencie2016quantum}, such as the 
von Neumann entropy, exhibit the usual linear growth at short times, 
followed by a volume-law scaling at asymptotically long times. 
%
%####################################################
\begin{figure}[t]
\includegraphics[width=.45\textwidth]{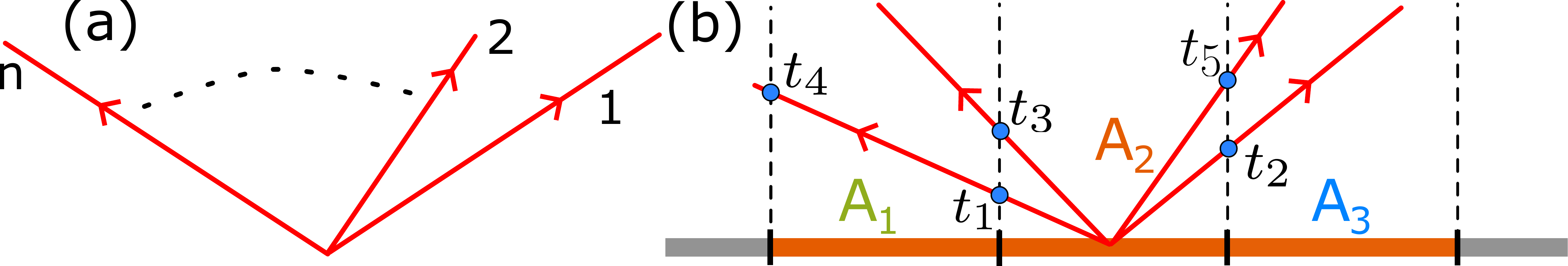}
\caption{Multiparticle entangled excitations in spin chains and 
	tripartite mutual information (TMI) dynamics. (a) Example of 
	an entangled multiplet formed by  $n$ different 
	quasiparticle species. (b) The TMI  (cf.~\eqref{eq:trip-def}) measures correlations 
	shared between the three intervals $A_j$ and between them and the rest.  
	Here we focus on three adjacent  intervals $A_j$, $j=1,2,3$ of equal 
	length $\ell$. Within the quasiparticle picture, only 
	multiplets that are \emph{shared} between all the intervals 
	$A_j$, and between the intervals and the rest contribute to the TMI. 
	For instance, in (b) we 
	show a quadruplet produced in $A_2$. The circles denote 
	the times at which a quasiparticle changes subsystem. 
	For $t_2\le t\le t_3$ the quadruplet is shared between all the subsystems, but not with 
	their complement $\bar A$. Thus, the TMI is zero for $0\le t\le t_3$. At $t_3$ the 
	leftmost particle leaves $A_1$ and the quadruplet starts to contribute to the TMI. 
}
\label{fig:cartoon}
\end{figure}
%####################################################
%

Here we show that the  tripartite mutual information is a much more revealing  
tool to highlight the presence of entangled multiplets and the concomitant breaking of the
relationship between entanglement  and thermodynamics. 
The TMI became popular~\cite{kitaev2006topological,levin2006detecting} as a witness of 
topological order, which is an intrinsically nonlocal quantum correlation. 
Let us consider a tripartition of a subsystem $A$ as $A=A_1\cup A_2\cup A_3$ (see Fig.~\ref{fig:cartoon}). 
The tripartite mutual information is defined as~\cite{cerf1998information} 
\begin{equation}
\label{eq:trip-def}
I_3:=I_2(A_1:A_2)+I_2(A_2:A_3)-I_2(A_2:A_1\cup A_3). 
\end{equation}
Here the mutual information $I_2$ measures the correlation between two intervals, and 
it is defined as 
\begin{equation}
\label{eq:mi-def}
I_2(A_i:A_j):=S(A_i)+S(A_j)-S(A_i\cup A_j), 
\end{equation}
where $S(A_j)$ is the von Neumann entropy of subsystem $A_j$. We consider only the 
case of three adjacent intervals of equal length $\ell$ (see Fig.~\ref{fig:cartoon}). 
The generalization to the case of disjoint intervals is straightforward.

The TMI was studied extensively~\cite{casini2009remarks} in free Quantum Field Theories (QFTs) 
(see also Ref.~\cite{agon2022tripartite} for recent results) at equilibrium. Interestingly, 
in holographic theories one can show~\cite{hayden2013holographic} that $I_3\le 0$, which means that 
the mutual information is monogamous. This suggests that correlations are genuinely 
quantum. Indeed, as it is clear from~\eqref{eq:trip-def}, a negative TMI reflects that the correlation 
shared between the three intervals is more than the sum of the pairwise correlations between them, and hence 
is quantum-delocalized. 
Some general results on the sign of the 
TMI in random states of few  qubits was presented in Ref.~\cite{rangamani2015entanglement}.  
It is challenging, however, to obtain the TMI in equilibrium and out-of-equilibrium 
quantum many-body systems~\cite{iyoda2018scrambling,seshadri2018tripartite,schnaack2019tripartite,pappalardi2018scrambling}. 
In out-of-equilibrium systems, a negative TMI is routinely used as a fingerprint of the so-called quantum information 
scrambling~\cite{hayden2007black,sekino2008fast,hosur2016chaos}, which is associated with chaotic 
dynamics. Chaotic systems lack a well-defined 
notion of quasiparticles, implying that the spreading of quantum information 
does not happen in a ``localized'' manner, for instance, via the propagation of entangled quasiparticles. 
As a result, quantum information is quickly dispersed in the global correlations. 
A ``weak'' form of scrambling  is present in integrable 
systems as well~\cite{alba2019quantum,modak2020entanglement}. 
A negative TMI was also linked with thermalization in CFTs with a gravity dual~\cite{balasubramanian2011thermalization,allais2012holographic}. The TMI received constant attention in 
generic CFTs~\cite{nie2019signature,kudler-flam2020quantum,kudler2020correlation}. 
Interestingly, it was shown~\cite{zabalo2020critical} that in the so-called ``minimal-cut'' picture 
for entanglement spreading~\cite{nahum2017quantum}, which 
is supposed to apply to chaotic systems, the TMI is always negative. 
This is supported by exact results in random local unitary circuits~\cite{bertini2020scrambling}, 
showing that  the TMI decreases linearly with time. 
Recently, it was shown that~\cite{maric2023universality,maric2023universalityin} 
in one-dimensional models the steady-state TMI after a quantum quench   
admits a field-theoretical interpretation. 
Finally,  in free-fermion models under continuous monitoring the TMI 
is negative at any time, and saturates to a negative value in 
the steady state~\cite{carollo2022entangled}.

Here we show that the TMI can be negative even after quenches in integrable spin chains that can be mapped onto 
free theories. Precisely, we consider quenches from low-entanglement initial states 
in the so-called $XX$ chain and in the quantum Ising 
chain with inhomogeneous transverse field. We focus on quenches that produce entangled multiplets of quasiparticles. 
For quenches that produce only entangled pairs, the TMI vanishes in the so-called hydrodynamic limit $t,\ell \to \infty$,
with fixed ratio $t/\ell$ (see Ref.~\cite{parez2022analytical} for 
a derivation for quenches  in the $XY$ chain);
this happens because the pairs can entangle only two intervals at a time. 
For the following, we should stress that all our results hold in the hydrodynamic limit. 
We show that, despite the presence of multiplets, it is possible to construct 
a quasiparticle picture for the TMI. First, only  multiplets that 
are shared between the three intervals $A_1,A_2,A_3$ and between the intervals and their complement $\bar A$ 
contribute to the TMI (as illustrated in  Fig.~\ref{fig:cartoon} (b)). This implies that only multiplets formed by 
$n>3$ quasiparticles give rise to nonzero TMI, as it was already shown in Ref.~\cite{carollo2022entangled} in 
a specific setting. 
For generic multiplets the TMI can be both positive and negative. 
For instance, we prove that 
for the ``classical'' multiplets considered in Ref.~\cite{bertini2018entanglement} the 
TMI is positive at all times. 
Oppositely, for quantum-correlated  multiplets (as in Ref.~\cite{bastianello2018spreading}) 
the TMI attains negative values during the dynamics, although it vanishes at 
asymptotic long times. 
Thus, a negative TMI is associated with the breaking of the relationship between entanglement and GGE. 
It is also intriguing to observe that the negative TMI at intermediate times 
reflects that correlations are nontrivially ``scrambled'' in the degrees of freedom of the multiplets. 

The manuscript is organized as follows. In section~\ref{sec:models} we introduce the $XX$ chain and the 
Ising chain with staggered magnetic field. In section~\ref{sec:genmethod} we discuss the general strategy to 
construct the quasiparticle picture for the TMI in the presence of generic multiplets. 
In particular, in section~\ref{sec:e-bertini} we focus 
on the states of Ref.~\cite{bertini2018entanglement}. In section~\ref{sec:equivalence} we show how the results of 
Ref.~\cite{bertini2018entanglement} fit into the framework of section~\ref{sec:genmethod}. 
In section~\ref{sec:positive_trip} we prove that the ``classically'' entangled multiplets discussed 
in Ref.~\cite{bertini2018entanglement} yield positive TMI at all times. 
In section~\ref{sec:negative_free} and section~\ref{sec:ising} 
we provide  examples of quenches that give negative TMI in the $XX$ chain and in the Ising chain, respectively. 
In section~\ref{sec:num} we benchmark our results against numerical data for the  
$XX$ chain and the Ising chain (in section~\ref{sec:xx_num} and section~\ref{sec:ising_checks}). We present our 
conclusions in section~\ref{sec:concl}. 
In Appendix~\ref{sec:app} we provide an \emph{ab initio} derivation of the quasiparticle picture for 
the single-interval von Neumann entropy for the same quench discussed in section~\ref{sec:negative_free}. 

%####################################################
\section{Models and out-of-equilibrium protocols}
\label{sec:models}

Here we consider the so-called $XX$ chain defined as 
\begin{equation}
	\label{eq:ham-xx}
	H=-J\sum_{i=1}^{L}(\sigma^x_i\sigma_{j+1}^x+\sigma_j^y\sigma_{j+1}^y)
	+h\sum_{i=1}^L\sigma_i^z. 
\end{equation}
Here $\sigma_j^{\alpha}$, with $\alpha=x,y,z$, are the Pauli matrices. We employ periodic 
boundary conditions setting $\sigma_{L+1}^\alpha=\sigma_1^\alpha$, and we fix $J=1$ and $h=0$. 

The $XX$ chain after a Jordan-Wigner transformation is transformed in 
a tight-binding fermionic chain as 
\begin{equation}
	\label{eq:freechain}
	H=- \sum_{i=1}^{L} (c_i^\dagger c_{i+1} + c_{i+1}^\dagger c_i),
\end{equation}
where $c^\dagger_i,c_i$ are standard fermionic operators acting at site $i$ of 
the chain, and obeying the standard fermionic anticoomutation relations 
$\{c_j,c^\dagger_l\}=\delta_{jl}$. 
The Hamiltonian can be easily diagonalized through a Fourier transform
\begin{equation}
\label{eq:free_ft}
c_k=\frac{1}{\sqrt{L}} 
\sum_{j=1}^L e^{ikj} c_j,\quad k=\frac{2\pi r}{L},\,\, r\in[1,L]. 
\end{equation}
We can rewrite~\eqref{eq:ham-xx} as 
\begin{equation}
\label{eq:freechain_diag}
H=\sum_k \varepsilon(k) c_k^\dagger c_k, \quad\mathrm{with}\,\, \varepsilon(k)=-2\cos(k), 
\end{equation}
where $\varepsilon(k)$ gives the single-particle dispersion of the fermions.

We also consider the inhomogeneous transverse field Ising chain, defined as 
\begin{equation}
\label{eq:ham-is}
H=J\sum_{i=1}^L\sigma_{i}^x\sigma_{i+1}^x+\sum_{i=1}^Lh_i\sigma_i^z. 
\end{equation}
Here the magnetic field $h_i$ is site-dependent. In particular, here we 
consider the case in which $h_j$ has a periodicity $n$, i.e., $h_{j}=h_{j+n}$. 
Again, we consider periodic boundary conditions for the spins. 
The inhomogeneous Ising Hamiltonian~\eqref{eq:ham-is} after the Jordan-Wigner transformation 
becomes 
\begin{equation}
\label{eq:ising-c}
H=\sum_{j=1}^L \Big[-\frac{1}{2} (c_{j}^\dagger c_{j+1}^\dagger + c_j^\dagger c_{j+1} 
+ \text{h.c.}) + h_j c_j^\dagger c_j \Big], 
\end{equation}
where $c_j,c_j^\dagger$ are fermionic annihilation and creation operators and again we assume $J=1$. 
We should remark that for the Ising chain the Jordan-Wigner transformation 
introduces some ambiguity in the boundary conditions for the fermions (see Ref.~\cite{calabrese2012quantum} 
and~\cite{calabrese2012quantumquench} for a discussion). For a generic 
global quantum quench these boundary conditions have no effect on the entanglement 
dynamics. Here we neglect them, choosing periodic boundary conditions 
also for the fermions.

For the following, it is crucial to observe that both the 
tight-binding~\eqref{eq:freechain} and the fermionic Ising chain~\eqref{eq:ising-c}  
can be reduced to the form 
\begin{equation}
\label{eq:diag-gen}
H=\int_{\mathcal B} \frac{dk}{2\pi} \sum_{j=1}^n \varepsilon_j(k) \eta_j^\dagger (k) 
\eta_j(k),
\end{equation}
The sum over $j$ is a sum over different species of quasiparticles, $\varepsilon_j(k)$ is 
the dispersion of the individual species, and ${\mathcal B}$ is a reduced 
Brillouin zone for the different species. In~\eqref{eq:diag-gen},  $\eta_j(k)$ is a 
fermionic operator with quasimomentum $k$ and of species $j$. Crucially, the choice of the 
different species of quasiparticles depends on the symmetry of the initial state, as we now 
discuss. 

Let us first consider the $XX$ chain, with the prequench initial states 
$|\Phi_{\{a_1,\dots, a_\nu\}}^\nu\rangle$ 
considered in Bertini et al.~\cite{bertini2018entanglement}. In the protocol of Ref.~\cite{bertini2018entanglement} 
the system is prepared in a state obtained from 
a unit cell of $\nu$ sites repeated $L/\nu$ times. Importantly, in 
each cell there is a single fermion that can be in  a generic  pure quantum state. 
Thus, the initial state $|\Phi^\nu_{\{a_1,\dots,a_\nu\}}\rangle$ is of the form 
\begin{equation}
\label{eq:free_states}
|\Phi^\nu_{\{a_1,\dots, a_\nu\}}
\rangle=\prod_{j=0}^{L/\nu-1}\Big(\sum_{m=1}^{\nu} a_m c_{\nu j +m}^\dagger\Big)
|0\rangle, 
\end{equation}
where the coefficients $a_m$ are arbitrary, and are normalized as $\sum_{m=1}^\nu |a_m|^2=1$. 
As it is clear from~\eqref{eq:free_states} the state $|\Phi^\nu\rangle$ is obtained by 
repeatition of a $\nu$-sites unit cell. Within the unit cell there is only one particle, 
which can be in a generic superposition state. 
The  states in~\eqref{eq:free_states} are Gaussian, as proved in Ref.~\cite{bertini2018entanglement}. 
In the following we refer to states of the form~\eqref{eq:free_states} as ``classically" entangled 
states (see section~\ref{sec:positive_trip}). 

Let us now observe that states of the form~\eqref{eq:free_states} have 
non-vanishing overlap (cf.~\cite{mazza2016overlap}) only with eigenstates of the $XX$ chain of the 
form~\cite{bertini2018entanglement} 
\begin{equation}
	\label{eq:free-generic}
	|\Psi_{k_1,\dots,k_N}\rangle = c_{k_1}^\dagger... c_{k_N}^\dagger |0\rangle, 
\end{equation}
with the conditions 
\begin{equation}
	\label{eq:constr}
	N=\frac{L}{\nu},\quad k_i-k_j\ne 0\,\mathrm{mod}\,\,\frac{2\pi}{\nu},\quad i,j=1,\dots,\frac{L}{\nu}. 
\end{equation}
Here the first constraint in~\eqref{eq:constr} takes into account that the 
number of fermions in the state~\eqref{eq:free_states} is $L/\nu$. 
The second condition ensures that only one quasimomentum (defined modulo $2\pi$) 
in the set ${\mathcal B}_\nu(k)$
\begin{equation}
	\label{eq:constr-2}
	{\mathcal B}_\nu(k)=\Big\{k, k+\frac{2\pi}{\nu}, ..., k+(\nu-1)\frac{2\pi}{\nu}\Big\}, 
\end{equation}
can appear in $|\Psi_{k_1,k_2,\dots,k_N}\rangle$. 
Violation of~\eqref{eq:constr-2} gives zero overlap. 

To enforce the constraint~\eqref{eq:constr}, we can restrict the Brillouin zone choosing  
$k\in (\pi-\frac{2\pi}{\nu},\pi]$. We can define new fermionic 
operators $\eta^\dagger_j(k)$ as 
\begin{equation}
\label{eq:freechain_redef}
\eta_j^\dagger(k)=
c^\dagger_{k-(j-1)2\pi/\nu}, \quad 
\varepsilon_j(k)=\varepsilon\Big(k-(j-1)\frac{2\pi}{\nu}\Big),
\end{equation}
where $\varepsilon(k)$ is the dispersion~\eqref{eq:freechain_diag}, 
and $j=1,\dots,\nu$ runs over the quasiparticles species. 
It is clear that by employing~\eqref{eq:freechain_redef} the Hamiltonian~\eqref{eq:freechain_diag}
becomes of the form~\eqref{eq:diag-gen}, with $n=\nu$ and the Brillouin 
zone ${\mathcal B}=(\pi-2\pi/\nu,\pi]$. Finally, we should remark that although 
Eq.~\eqref{eq:diag-gen} is only a rewriting of~\eqref{eq:freechain_diag}, it has 
the advantage that it is compatible with the translation invariance of the initial state. 
Moreover,  as we will discuss in section~\ref{sec:genmethod}, the correlations, and hence entanglement, 
generated by the out-of-equilibrium dynamics can be conveniently encoded in the correlation 
between the species operators $\eta_j$. 

For the free fermion chain \ref{eq:freechain}, we also consider the initial 
state $|\Phi_0\rangle$ defined as 
\begin{equation}
\label{eq:ex_init}
|\Phi_0\rangle= \prod_{j=0}^{L/4-1}(c_{4 j +1}^\dagger c_{4j+2}^\dagger)
|0\rangle;
\end{equation}
This initial state does not fall into the class of initial states 
considered in Ref.~\cite{bertini2018entanglement}. Indeed, although the 
state~\eqref{eq:ex_init} is constructed from the repetition of a 
four-site unit cell $|1100\rangle$, the unit cell contains more than 
one particle. Still, the state~\eqref{eq:ex_init} is Gaussian, 
as Wick's theorem applies. The quench from the state~\eqref{eq:ex_init} was also 
studied in Ref.~\cite{najafi2018light}. 
The state has nonzero overlap with the $XX$ chain eigenstates $|\Psi_{k_1,k_2,\dots,k_N}\rangle$ 
satisfying the constraint that \textit{at most} three quasimomenta (defined modulo $2\pi$) 
of ${\mathcal B}_4(k)$ (cf.~\eqref{eq:constr-2}) appear. 
We anticipate that, in contrast with the states~\eqref{eq:free_states}, 
the state~\eqref{eq:ex_init} gives rise to a negative TMI. 

Let us now discuss the case of the inhomogeneous Ising chain. 
In our out-of-equilibrium protocol, the system is initially prepared in the 
initial state $|\Psi_0\rangle$, which  
is the ground state of the Hamiltonian~\eqref{eq:ham-is} 
with initial magnetic field $h^{0}_j=(h^0_1,h^0_2,\dots,h^0_n)$. 
The magnetic field is then instantly changed to  (cf.~\eqref{eq:ham-is})
$h_j=(h_1,h_2,\dots, h_n)$~\cite{bastianello2018spreading}. 

First, let us diagonalize the post-quench Hamiltonian~\eqref{eq:ham-is} defining the 
quasiparticle excitations $\eta_j(k)$. Following Ref.~\cite{bastianello2018spreading} 
we restrict the Brillouin zone to $[0,\pi/n)$. In the limit $L\to\infty$, 
we can rewrite the Ising Hamiltonian~\eqref{eq:ham-is} as 
\begin{equation}
\label{eq:h-supp}
H=\int_0^{\pi/n}\frac{dk}{2\pi}C^\dagger(k){\mathcal H}_k C(k), 
\end{equation}
where $C^\dagger$ is the $2n$-dimensional vector of the Fourier 
transform of the original fermions $c_j$ (cf.~\eqref{eq:ising-c}) 
defined as 
\begin{equation}
\label{eq:CDdef}
C^\dagger(k)=\big(c^\dagger_k,\dots, c^\dagger_{k+(n-1)\pi/n}, 
c_{-k},\dots, c_{-k-(n-1)\pi/n}\big). 
\end{equation}
In~\eqref{eq:h-supp}, ${\mathcal H}_k$ is a $2n\times 2n$ matrix encoding 
the Hamiltonian~\eqref{eq:ising-c}. To proceed, we can diagonalize 
$h_k$  by defining new fermions $D_{h}(k)$ as 
\begin{equation}
\label{eq:CDdef-1}
D_{h}^\dagger=\big(d_1^\dagger(k),\dots,
d_n^\dagger(k), d_1(-k),\dots, d_n(-k)\big).
\end{equation}
The fermions $d(k)$ are defined via the relationship 
\begin{equation}
\label{eq:CDunit}
C(k)=U_{h}(k) D_{h}(k), 
\end{equation}
with $C(k)$ as in~\eqref{eq:CDdef}. 
In~\eqref{eq:CDunit}, $U_h$ is a $2n\times 2n$ unitary matrix, which is 
determined by requiring that in terms of $d_j(k)$ and $d_j(-k)$ the free-fermion 
Hamiltonian~\eqref{eq:ising-c} becomes diagonal. For generic $n$, $U_h$ has to be 
determined numerically. For $n=1$ one recovers the standard Bogoliubov 
transformation~\cite{calabrese2012quantum}. Now Eq.~\eqref{eq:ising-c} becomes 
\begin{equation}
\label{eq:ising_diag}
H=\int_0^{\pi/n} \frac{dk}{2\pi} \sum_{j=1}^n 
\varepsilon^h_j(k) (d_j^\dagger(k)d_j(k)-d_j(-k)d_j^\dagger(-k) ). 
\end{equation}
In~\eqref{eq:ising_diag} $\pm\varepsilon^h_j(k)$, with $\varepsilon_j^h(k)\ge0$ are the 
eigenvalues of $h_k$ (cf.~\eqref{eq:h-supp}), and form  the single-particle 
dispersion. The ground state of~\eqref{eq:ising_diag} is annihilated by all 
the operators $d(\pm k)$. A similar procedure  allows to diagonalize  the pre-quench 
Hamiltonian, with different sets of operators $D_{h^{0}}(k)$.  
The latter are obtained from the original fermions $c_k$ via a 
different unitary transformations $U_{h^{\scriptscriptstyle 0}}(k)$. 

Crucially, since the fermionic operators $c_k$  (cf.~\eqref{eq:CDdef}) 
are the same before and after the quench, the operators diagonalizing the pre-quench 
and post-quench Hamiltonians are linked by a unitary transformation as 
\begin{equation}
	\label{eq:unitdiag}
	D_{h}(k)=W(k) D_{h^{0}}(k),\quad W=U_h^{-1} U_{h^0}. 
\end{equation}
Let us now identify
\begin{equation}
\label{eq:ising_redef}
\eta_j(k)=\left\{\begin{array}{cc}
		d_j(k) & j\in[1,n]\\\\
		d^\dagger_{j-n}(-k) & j\in[n+1,2n]
\end{array}\right.
\end{equation}
After employing the definitions in~\eqref{eq:ising_redef} the inhomogeneous Ising 
chain becomes of the form~\eqref{eq:diag-gen}. 
Again, unlike the homogeneous Ising 
chain~\cite{calabrese2012quantum,calabrese2012quantumquench}, for the inhomogeneous one it is not possible 
in general to 
obtain analytically the single-particle dispersion $\varepsilon_i(k)$ and the matrices 
$W(k)$ (cf.~\eqref{eq:unitdiag}) and $U_h$ (cf.~\eqref{eq:CDunit}). However, as they are 
$2n\times 2n$ matrices, they can be obtained numerically with modest computational cost. 

To determine the dynamics of the TMI it is necessary to compute the correlation functions 
(see section~\ref{sec:genmethod}) 
\begin{equation}
	\label{eq:ck}
	{\mathcal C}(k) = \langle 0|
	\left(\begin{array}{c}
			\eta_1\\
			\vdots\\
			\eta_{2n}\\
			\eta_1^{\dagger}\\
			\vdots\\
			\eta_{2n}^\dagger
	\end{array}\right)
	\left(\begin{array}{cccccc}
			\eta_1^\dagger & \cdots&\eta_{2n}^\dagger & \eta_1 &\cdots &\eta_{2n}
	\end{array}\right)
	|0\rangle, 
\end{equation}
where $|0\rangle$ is the ground state of the Ising chain with magnetic field 
$h^0$, and $\eta_j$ are the operators that diagonalize the Ising chain with magnetic field 
$h$. It is straightforward to compute the correlator~\eqref{eq:ck} by first using 
Eq.~\eqref{eq:unitdiag}, and then by using that the operators $\eta^{\scriptscriptstyle(0)}_j$ of the initial 
Ising chain annihilate the ground state. Hence we obtain 
\begin{equation}
	{\mathcal C}(k)=\left(\begin{array}{cc}
			W(k) & 0\\
			0 & W^*(k)
	\end{array}\right){\mathcal C}^{(0)}
	\left(\begin{array}{cc}
			W^\dagger (k) & 0\\
			0 & W^T(k)
	\end{array}\right), 
\end{equation}
where $W(k)$ is defined in~\eqref{eq:unitdiag}, and ${\mathcal C}^{\scriptscriptstyle (0)}$ is a $4n\times 4n$ diagonal 
matrix ${\mathcal C}^{\scriptscriptstyle(0)}_{ij}=\delta_{ij}$ for $i\in[1,n]\cup[3n+1,4n]$, and zero 
otherwise. The matrix ${\mathcal C}^{\scriptscriptstyle(0)}$ is the correlation of the pre-quench 
operators $\eta_j^{\scriptscriptstyle(0)}$ calculated over the initial state. 

%####################################################
\section{Quasiparticle picture in the presence of entangled multiplets}
\label{sec:genmethod}
	
Here we show how to determine the quasiparticle picture for the TMI in the presence of 
entangled multiparticle excitations. We start from the framework developed in 
Ref.~\cite{bastianello2018spreading} (see also~\cite{bertini2018entanglementevolution}). 

Let us also assume that the Hamiltonian governing the post-quench dynamics can be diagonalized by a set of 
operators $\eta_1^\dagger (k), \eta_2^\dagger (k), ... \eta_n^\dagger (k)$ as in~\eqref{eq:diag-gen}. 
Let us also assume that the two-point correlation function of the fermionic operators $\eta_j(k)$ 
calculated on the initial state is block-diagonal as 
\begin{equation}
\label{eq:blockd-2pt}
\mathcal{C}(k,p):=\langle \Psi_0 | \Gamma (k) \Gamma^\dagger (p) 
| \Psi_0 \rangle \propto \delta_{k,p},
\end{equation}
where
\begin{equation}
	\Gamma^\dagger (k) :=(\eta_1^\dagger (k), ..., \eta_n^\dagger (k), \eta_1 (k), ...,\eta_n (k) ). 
\end{equation}
It is straightforward to check that all the protocols we introduced in~\ref{sec:models}
satisfy this requirement. 
We now consider the correlation matrix ${\mathcal C}(k)$ at fixed quasimomentum $k$ but in the 
space of species. Specifically, we can write 
\begin{equation}
\label{eq:2pt_Cblock}
{\mathcal C}(k):=
\langle \Psi_0 | \Gamma (k) \Gamma^\dagger (k) | \Psi_0 \rangle=\begin{pmatrix}
\mathbb{1}-G^T(k) & F(k) \\ 
F^\dagger(k)& G(k)
\end{pmatrix},
\end{equation}
where  the correlators $G_{ij}(k)$ and $F_{ij}(k)$ are $n\times n$ 
matrices defined as 
\begin{align}
	\label{eq:Gij}
	& G_{ij}(k):=\langle\Psi_0|\eta^\dagger_i(k)\eta_j(k)|\Psi_0\rangle\\
	\label{eq:Fij}
	& F_{ij}(k):=\langle\Psi_0|\eta_i(k)\eta_j(k)|\Psi_0\rangle. 
\end{align}
Notice that since $G_{ij}$ and $F_{ij}$ are not diagonal, they encode nontrivial correlations 
between the different species of quasiparticles. 
The von Neumann entropy of a subregion $A$ and generic entanglement-related quantities are 
obtained from~\eqref{eq:2pt_Cblock} (see~\cite{peschel2009reduced}). Indeed, by taking 
the inverse Fourier transform 
of ${\mathcal C}(k)$ one obtains the fermionic correlation function $\widetilde{\mathcal C}_{nm}$ 
in real space. From that, the von Neumann entropy is written as~\cite{peschel2009reduced}  
\begin{equation}
	S_A=-\mathrm{Tr}\,\,\widetilde {\mathcal C}_A\ln \big(\widetilde {\mathcal C}_A \big), 
\end{equation}
where ${\mathcal C}_A$ is the correlation matrix restricted to $A$, i.e., with $n,m\in A$.

Before proceeding, let us observe that 
for any fixed $k$, the correlation matrix ${\mathcal C}(k)$ (cf.~\eqref{eq:2pt_Cblock}) 
is the covariance matrix of a Gaussian pure state. This stems from the fact that the 
${\mathcal C}$ of the full system can have only the eigenvalues $0,1$ because the system 
is in a pure state and ${\mathcal C}$ has a block structure in quasimomentum 
space, implying that each block with fixed $k$ can only have eigenvalues $0,1$. 

The correlation matrix~\eqref{eq:2pt_Cblock} is the main ingredient  to build a quasiparticle 
picture in the presence of entangled multiparticle excitations~\cite{bastianello2018spreading}. 
In the quasiparticle picture, at time $t=0$,
at each point in space a multiplet is produced, with arbitrary quasimomentum $k$.  
At later times  the quasiparticles forming the multiplet spread, 
each quasiparticle species propagating with group velocity 
$v_i(k)=d \varepsilon_i(k)/dk$, with $\varepsilon_i(k)$ the single-particle energies 
in~\eqref{eq:diag-gen}. The growth of the von Neumann entropy 
of a subsystem $A$ is attributed to the quasiparticles of the same 
multiplet that are shared between $A$ and the rest.

We now determine the contribution at time $t$ to the entanglement entropy $S_A$ 
of a region $A$ (see Fig.~\ref{fig:cartoon}) of an entangled multiplet with quasimomentum 
$k$. Let us consider the situation in which at time $t$ only a subset  $m$ of the $n$ 
quasiparticles forming the multiplet is in $A$, the remaining ones being in the complement of 
$A$. 
The quasiparticles in $A$  correspond to some operators $\eta_{i_1},\eta_{i_2},\dots,\eta_{i_m}$, 
where $1\le i_p\le n$. We introduce the matrix ${\mathcal C}_A(k,{\mathcal Q}_A)$, with 
${\mathcal Q}_A=\{i_p\}_{p=1}^m$ as the correlation matrix ${\mathcal C}(k)$ 
(cf.~\eqref{eq:2pt_Cblock}) in which we restrict the row and column indices of 
$G_{ij}$ and $F_{ij}$  (cf.~\eqref{eq:Gij} 
and~\eqref{eq:Fij})  to the subset ${\mathcal Q}_A$. Finally, the 
contribution of this configuration to the entanglement entropy is 
\begin{equation}
	\label{eq:s_A}
	s(k,{{\mathcal Q}_A})=-\mathrm{Tr}\,\,{\mathcal C}_A\ln({\mathcal C}_A), 
\end{equation}
where the trace is over the $2m\times 2m$ matrix ${\mathcal C}_A(k)$. Again, in~\eqref{eq:s_A} 
${\mathcal Q}_A$ are the indices of the quasiparticles that are in $A$. 
Notice that $s(k,{\mathcal Q}_A)=s(k,{\mathcal Q}_{\bar A})$. This is due to 
the fact that ${\mathcal C}(k)$ defines a Gaussian pure state. 
Finally, the entanglement entropy $S_A$ is obtained as 
\begin{equation}
	\label{eq:SA-fin}
	S_A=\int_{\mathcal B}\frac{dk}{2\pi}\sum_{{\mathcal Q}_A}
	{\mathcal D}(k,{\mathcal Q}_A,\ell,t) s(k,{\mathcal Q}_A). 
\end{equation}
Here the sum is over all the possible ways of distributing the 
quasiparticles forming the entangled multiplet between $A$ and its 
complement. In~\eqref{eq:SA-fin} ${\mathcal D}(k,{\mathcal Q}_A,\ell,t)$ 
is a kinematic factor that counts the number of entangled multiplets 
with fixed $k$ created at $t=0$ and that at time $t$ give rise to the 
configuration ${\mathcal Q}_A$. The factor ${\mathcal D}(k,{\mathcal Q}_A,\ell,t)$ 
depends on time and on the length $\ell$ of $A$. Moreover, it depends on $k$ 
through the velocities $\varepsilon_j'(k)$ of the quasiparticles. 

Finally, we should stress that although 
the presence of entangled multiplets does not invalidate the quasiparticle 
picture for entanglement spreading, the entanglement content $s(k,{\mathcal Q}_A)$ of the quasiparticles 
is not directly related to the thermodynamic entropy of the GGE that describes the 
steady state, unlike the case in which only entangled pairs of quasiparticles are 
produced after the quench~\cite{calabrese-2005,fagotti2008evolution,alba2017entanglement}. 
In particular, the entanglement content does not depend only on the diagonal correlations in~\eqref{eq:Gij} 
that represent the root densities of the excitations 
\begin{equation}
	\label{eq:rhoj}
	\rho_j(k):=\langle\Psi_0|\eta_j^\dagger(k)\eta_j(k)|\Psi_0\rangle,
\end{equation}
while the GGE contains information only about these densities ~\cite{bastianello2018spreading}. 
Let us briefly discuss the relationship between entanglement entropy and GGE thermodynamic entropy 
in quenches in free-fermion systems~\cite{alba2018entanglement}. First, in the limit $t\to\infty$ after 
a quench from typical initial states, it is well established that local observables reach a stationary value, which 
is  describable via a statistical ensemble. Since free-fermion models are integrable, this is not the usual Gibbs ensemble~\cite{vidmar2016generalized}. 
The correct ensemble is the so-called Generalized Gibbs Ensemble (GGE), which can be fully determined by 
the occupations $\rho_j(k)$ in~\eqref{eq:rhoj}. 
Importantly, in the thermodynamic limit $L\to\infty$ 
there is an exponentially diverging number of \emph{microscopic} eigenstates of the model that give 
rise to the same GGE, or, equivalently, to the same occupations $\rho_j(k)$.  
In the thermodynamic limit, expectation values of local observables over these eigenstates become the 
same. The number of microscopic eigenstates that give rise to the same thermodynamic \emph{macrostate} 
is given by the so-called Yang-Yang entropy~\cite{alba2021generalized} $S_{YY}$ defined as 
\begin{equation}
\label{eq:yy}
	S_{YY}:=L\sum_j\int_{-\pi}^\pi \frac{dk}{2\pi}s_j^{YY}(j,k), 
\end{equation}
where~\cite{alba2021generalized} 
\begin{equation}
	s_j^{YY}=\rho_j(k)\ln(\rho_j(k))
	+(1-\rho_j(k))\ln(1-\rho_j(k)). 
\end{equation}
Nevertheless,
in all the cases we take into account, the relationship between the entanglement content and the thermodynamic
entropy of the GGE is recovered in the limit $t/\ell \to \infty$, because in this limit the particles of a multiplet are too
far from each other, and only one of them is in $A$. Having $F_{ij}=0$ in all the cases we 
consider (see below), the von Neumann entropy depends only on the root densities~\eqref{eq:rhoj} and, precisely, 
reduces to the thermodynamic entropy~\eqref{eq:yy} of the GGE. 

To conclude, let us illustrate the formalism for the case of the quench from the 
N\'eel state. The N\'eel state 
$|101010\cdots\rangle$ corresponds to $n=\nu=2$ in~\eqref{eq:free_states} and to 
$a_1=1$ and $a_2=0$. Now, we have $\eta^\dagger_1=c^\dagger_k$ (cf.~\eqref{eq:diag-gen} and~\eqref{eq:freechain_redef}) and $\eta_2^\dagger=c^\dagger_{k-\pi}$, with $k\in (0,\pi]$. 
The pairing terms $F_{ij}$ (cf.~\eqref{eq:Fij}) are identically zero for the N\'eel quench. 
Moreover, $G_{jl}=1/2\;\delta_{jl}$ (cf.~\eqref{eq:Gij}) is diagonal and 
independent of $k$. Now, it is clear that there are only two ways of distributing the members of the pair between $A$ and the
complement. Precisely, one has ${\mathcal Q}_A(k)=\{1\}$ or ${\mathcal Q}_A(k)=\{2\}$ 
(cf.~\eqref{eq:SA-fin}). Notice that the energy of the two species of quasiparticles is  
$\varepsilon_1(k)=\varepsilon(k)$ and $\varepsilon_2(k)=\varepsilon(k-\pi)$, 
with $\varepsilon$ defined in~\eqref{eq:freechain_diag}. 
This implies that $v_1=\varepsilon'_1(k)=-v_2$. 
The kinematic function ${\mathcal D}(k,{\mathcal Q}_A,\ell,t)$ (cf.~\eqref{eq:SA-fin}) 
counts the number of pairs that are in the 
configuration ${\mathcal Q}_A$ at time $t$. It is clear that 
for ${\mathcal Q}_A=\{1\}$, one has that ${\mathcal D}(k,\{1\},\ell,t)$ 
takes contribution from the pairs created on the region near the 
left edge, which gives a contribution $\min(2v_1(k)t,\ell)$. 
The contribution of species $2$ is the same. Finally, 
it is clear that $s(k,\{1\})=s(k,\{2\})=\ln(2)$. This implies that 
\begin{equation}
	S_A=2\int_0^\pi \frac{dk}{2\pi} \min(2v_1(k)t,\ell)\ln(2),
\end{equation}
which is the well-known quasiparticle picture for the von Neumann entropy after the quench from the 
N\'eel state in the $XX$ chain~\cite{alba2018entanglement}. 

%####################################################
\subsection{An example of ``classically'' entangled multiplets: the states of Bertini et al.}
\label{sec:e-bertini}

In the last section we showed that the presence of nontrivially entangled multiplets of 
excitations implies that the dynamics of the von Neumann entropy cannot be always described in 
terms of the densities of excitations $\rho_j(k)$ (cf.~\eqref{eq:rhoj}). Still, as it has 
been pointed out in Ref.~\cite{bertini2018entanglement} (see also~\cite{bastianello2018spreading}) 
the out-of-equilibrium dynamics  
starting from the states $|\Phi_{\{a_1,\dots,a_\nu\}}^\nu\rangle$ (cf.~\eqref{eq:free_states}) 
in the $XX$ chain gives 
rise to ``classically'' entangled multiplets. As we will show in section~\ref{sec:positive_trip}, 
this implies that the TMI is positive  at all times. 

Let us now review the quasiparticle picture for the von Neumann entropy for quenches starting 
from the states $|\Phi^\nu_{a_1,a_2,\dots,a_\nu}\rangle$ (cf.~\eqref{eq:free_states}). 
Crucially, for this class of states the contribution of the entangled multiplets to the 
entropies is obtained in terms of the densities of the quasiparticles 
$\rho_j(k)$, which are defined as 
\begin{equation}
	\label{eq:bertini-rho}
	\rho_j(k)=\langle\Phi^\nu|\eta_j^\dagger(k) \eta_j(k)|\Phi^\nu \rangle, 
\end{equation}
where $\eta_j(k)$ are defined in~\eqref{eq:freechain_redef}. 
A structure similar to the one outlined in~\ref{sec:genmethod} emerges. 

The initial state acts as a source of entangled multiplets of one 
particle and $\nu-1$ holes. However, in contrast with the general picture 
of section~\ref{sec:genmethod}, the contribution of these multiplets is entirely 
written in terms of $\rho_j(k)$. Again, at a generic time $t$ we can consider 
the situation in which only a subset of the quasiparticles forming the multiplet 
is in $A$. Let us consider the case with $m$ quasiparticles $\eta_j$ 
with $j\in {\mathcal Q}_A$ in $A$. Let us define $\rho_{\mathrm{in}}(k)$ 
as 
\begin{equation}
	\label{eq:rho-in}
	\rho_{\mathrm{in}}(k)= 
	\sum_{j\in{\mathcal Q}_A}\rho_j (k). 
\end{equation}
The contribution of this configuration 
to the entanglement between $A$ and the rest is~\cite{bertini2018entanglement}  
\begin{equation}
	\label{eq:e-bertini}
	s(k,{\mathcal Q}_A)=-\rho_{\mathrm{in}}\ln(\rho_{\mathrm{in}})-
	(1-\rho_{\mathrm{in}})\ln(1-\rho_{\mathrm{in}}). 
\end{equation}
Notice that since the state of the system is pure, if all the quasiparticles are in 
$A$  one has $s(k,{\mathcal Q}_A)=0$. This means that 
$\sum_j\rho_j=1$. This constraint automatically implies that 
$s(k,{\mathcal Q})_A=s(k,{\mathcal Q}_{\bar A})$. 

From~\eqref{eq:e-bertini}, we obtain the entropy $S_A$ as 
\begin{equation}
	\label{eq:ent-bertini}
	S_A=\int_{\pi(1-2/\nu)}^{\pi}\frac{dk}{2\pi}\sum_{{\mathcal Q}_A}{\mathcal D}(k,{\mathcal Q}_A,\ell,t)s(k,{\mathcal Q}_A). 
\end{equation}
Here $s(k,{\mathcal Q}_A)$ is the entanglement content due to the 
configuration with the quasiparticles in ${\mathcal Q}_A$ being in $A$ and it 
is given in~\eqref{eq:e-bertini}. 
In~\eqref{eq:ent-bertini} the kinematic term ${\mathcal D}(k,{\mathcal Q}_A,\ell,t)$ 
counts the number of multiplets with fixed $k$ that are in ${\mathcal Q}_A$.

%####################################################
\subsection{Equivalence with the general method}
\label{sec:equivalence}

Let us show that the approach of Ref.~\cite{bertini2018entanglement} outlined in section~\ref{sec:e-bertini} 
corresponds to a particular case of the framework introduced in section~\ref{sec:genmethod}. 
We first observe that for the states $|\Phi^\nu_{\{a_1,\dots,a_\nu\}}\rangle$ 
one has that $F_{ij}=0$ (cf.~\eqref{eq:Fij}). Moreover, we have that 
\begin{equation}
	\langle \Phi^\nu | c_k^\dagger  c_{k'} |\Phi^\nu\rangle \neq 0\,
	\quad\text{only if} 
\,\,\nu (k-k')=0\,[\text{mod} \, 2\pi], 
\end{equation}
which follows from the $\nu$-site translation invariance. Now, Eq.~\eqref{eq:2pt_Cblock} 
is block diagonal as 
\begin{equation}
		\label{eq:freechain_2ptmat}
		{\mathcal C}(k)=\begin{pmatrix}
			\mathbb{1}-G^T(k) & 0 \\ 
		0& G(k)
	\end{pmatrix}. 
\end{equation}
Following the strategy of section~\ref{sec:genmethod}, we have to determine the 
entanglement entropy  associated to a partition  ${\mathcal Q}_A$ of the $\nu$ quasiparticles 
forming the entanglement multiplet. This is  
given by~\eqref{eq:s_A}. It is straightforward to show that~\eqref{eq:s_A} becomes  the 
same as~\eqref{eq:ent-bertini} provided that 
$G$ (cf.~\eqref{eq:freechain_2ptmat}) has rank one. Indeed, if the 
rank of $G$ is one, any submatrix $G_A(k)$ 
will have rank at most one. This means that its nonzero eigenvalue is $\mathrm{Tr}(G_A(k))$, 
with 
\begin{equation}
	\label{eq:trace}	
\mathrm{Tr}(G_A)=
\sum_{j\in {\mathcal Q}_A}\langle\Phi^\nu|\eta_j^\dagger\eta_j|\Phi^\nu\rangle=
\rho_{\mathrm{in}}(k), 
\end{equation}
where $\rho_{\mathrm{in}}$ is defined in~\eqref{eq:rho-in}. Finally, for a 
given set ${\mathcal Q}_A$ of quasiparticles in $A$,  
one obtains that the contribution $s(k,{\mathcal Q}_A)$ to the von Neumann entropy 
is $s(k,{\mathcal Q}_A)=-\mathrm{Tr}\,{\mathcal C}_A\ln({\mathcal C}_A)$ (cf.~\eqref{eq:freechain_2ptmat}), 
and by using~\eqref{eq:trace}, it coincides with~\eqref{eq:e-bertini}.

To conclude, we have to show that $G(k)$ for the generic state $|\Phi^\nu\rangle$ 
has rank one. By using the definition of $\eta_j(k)$ (cf.~\eqref{eq:freechain_redef}), 
we obtain 
\begin{multline}
\label{eq:Gmat_rk1_calc}
G_{jl}(k)
=\langle\Phi^\nu|\eta_j^\dagger(k)\eta_l(k)|\Phi^\nu\rangle\\
=\frac{1}{L}\sum_{m,n=1}^{L} e^{-i (k-(j-1)2\pi/\nu)m+i(k-(l-1)2\pi/\nu)n} 
\langle c_m^\dagger c_n \rangle
\\
=\frac{1}{\nu} \sum_{m,n=1}^{\nu} e^{-i (k-(j-1)2\pi/\nu)m} e^{i(k-(l-1)2\pi/\nu)n}\langle 
c_m^\dagger c_n \rangle\\
=\frac{1}{\nu} \sum_{m,n=1}^\nu e^{-i (k-(j-1)2\pi/\nu)m} a_m^*
 e^{i(k-(l-1)2\pi/\nu)n} a_n, 
\end{multline}
where we defined $\langle c^\dagger_m c_n\rangle:=\langle\Phi^\nu|c^\dagger_m c_n|\Phi^\nu\rangle$. 
In the second row in~\eqref{eq:Gmat_rk1_calc} we exploited translation invariance. The 
coefficients $a_j$ are defined in~\eqref{eq:free_states}. Now, it is clear that $G_{jl}$ 
has rank one for any $a_j$ because it is an outer product of two vectors.

%####################################################
\section{Classically-entangled multiplets  yield  non-negative TMI}
\label{sec:positive_trip}

\begin{figure}[t]
\includegraphics[width=.4\textwidth]{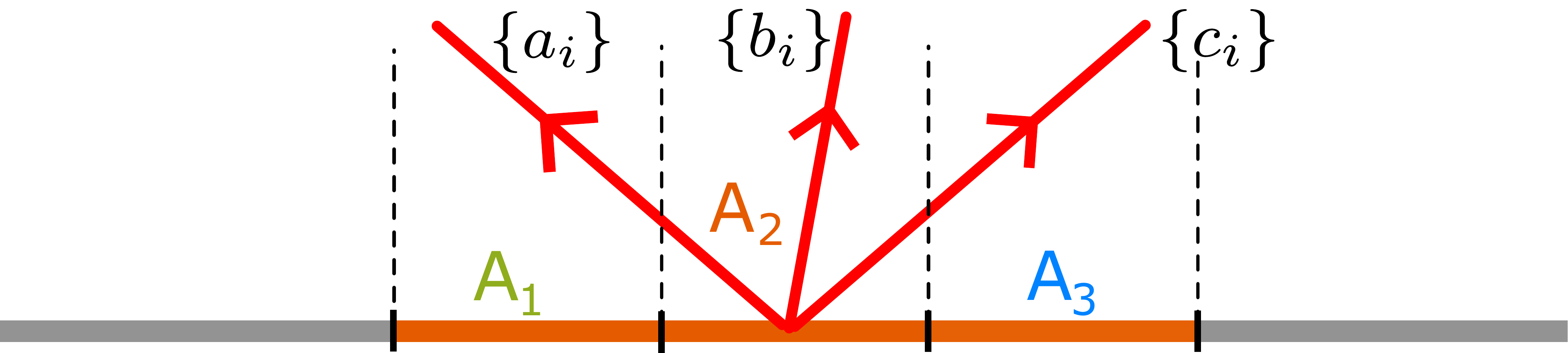}
\caption{ A typical entangled multiplet contributing to the dynamics 
 of the tripartite mutual information (TMI). 
 The entangled multiplet is created at $t=0$ in subsystem $A_2$, and 
 it consists of $n$ quasiparticles with labels ${\mathcal Q}=\{1,2,\dots,n\}$. 
 Here we denote as $\{a_i\}\subseteq{\mathcal Q}$ the quasiparticles in $A_1$. Similarly, 
 we define $\{b_i\}$ and $\{c_i\}$ as the 
 quasiparticles in $A_2$ and $A_3$, respectively. 
 The entanglement entropy of the  interval $A_1$ is obtained from 
 the restricted correlation matrix ${\mathcal C}_{A_1}(k)$ (cf.~\eqref{eq:2pt_Cblock}) 
 whose $F$ and $G$ blocks have row and column indices in $\{a_i\}$. 
}
\label{fig:cartoon-2}
\end{figure}

We now show that for all the quenches from the ``classically''  entangled 
states~\eqref{eq:free_states}, the tripartite mutual information (TMI) between 
three generic intervals is always non-negative in the 
hydrodynamic limit. Here for the sake of simplicity we consider the case 
of three equal adjacent intervals of length $\ell$ (see Fig.~\ref{fig:cartoon-2}). The hydrodynamic limit 
is defined as $\ell,t\to\infty$ with their ratio $t/\ell$ fixed. 

Given a generic entangled multiplet formed 
by $n$ quasiparticles, to build the quasiparticle picture for the TMI, we 
have to first identify the different 
ways of distributing the quasiparticles among the 
three subsystems. Let us denote by $\{a_i\}$, with $1\le a_i\le n$  
the set of indices identifying the quasiparticles that at a generic time 
$t$ after the quench are within $A_1$. Similarly, we can introduce $\{b_i\}$ and 
$\{c_i\}$ as the indices of the quasiparticles in $A_2$ and $A_3$, 
respectively (see Figure~\ref{fig:cartoon-2}). Notice that in general $\{a_i\}\cup \{b_i\}\cup\{c_i\}$ is 
not the full multiplet. Indeed, as it will be clear in the following, 
for the configuration to contribute to the 
TMI the multiplet has to be shared also with the complement of $A=A_1\cup A_2\cup A_3$. 

We can define the contribution $\tau_3$ of the quasiparticles to $I_3$ as 
\begin{multline}
\label{eq:trip_infinitesimal}
\tau_3(k,\{a_i\},\{b_i\},\{c_i\})=\\
s_{\{a_i\}\cup\{b_i\}\cup\{c_i\}}(k)-s_{\{a_i\}\cup\{b_i\}}(k)-
	s_{\{a_i\}\cup\{c_i\}}(k)\\
	-s_{\{b_i\}\cup\{c_i\}}(k)
	+s_{\{a_i\}}(k)+s_{\{b_i\}}(k)+s_{\{c_i\}}(k). 
\end{multline}
Precisely, the TMI is given as 
\begin{equation}
	I_3(t)=
	\sum_{\{a_i\},\{b_i\},\{c_i\}}\int_{\pi-\frac{2\pi}{\nu}}^{\pi} \frac{dk}{2\pi} \tau_3(k) {\mathcal D}(k,\ell,t), 
\end{equation}
where the sum is over the ways of distributing the quasiparticles forming the 
mutliplet among the three subsystems, and ${\mathcal D}(k,\ell,t)$ is a kinematic factor 
that describes the propagation of the quasiparticles forming the multiplet. 
For the von Neumann entropy and for the quenches that produce entangled pairs 
one has that ${\mathcal D}(k,\ell,t)=\min(2v(k)t,\ell)$. 
In Eq.~\eqref{eq:trip_infinitesimal} $s_{X}$ is the contribution to the von Neumann entropy
due to the quasiparticles $X$ being within the subsystem, and the remaining ones  
outside of it. 
In~\eqref{eq:trip_infinitesimal}, $s_X$ is obtained as the entropy of the 
reduced correlation matrix ${\mathcal C}_X(k)$ (cf.~\eqref{eq:s_A}). ${\mathcal C}_X$ 
is obtained from ${\mathcal C}(k)$ (cf.~\eqref{eq:2pt_Cblock}) by selecting the rows 
and columns in $X$. 
In~\eqref{eq:trip_infinitesimal} the first contribution is associated with 
the last  term in~\eqref{eq:trip-def}, i.e., with the entropy of $A_1\cup A_2\cup A_3$. 

Let us now discuss some constraints on $\{a_i,b_i,c_i\}$ to ensure a nonzero 
contribution to $\tau_3$. 
First, configurations without at least a quasiparticle in each of the three intervals $A_1,A_2,A_3$ 
give $\tau_3=0$. Indeed, without loss of generality we can assume that  
$\{a_i\}=\emptyset$, i.e., there are no quasiparticles in $A_1$ (see Fig.~\ref{fig:cartoon-2}). 
Then, from~\eqref{eq:trip_infinitesimal}, and using that $s_\emptyset(k)=0$, 
we have $\tau_3=0$.

%####################################################
%####################################################

An important consequence is that  one has nonzero $\tau_3$ only for $n\ge 3$, i.e., when 
triplets or larger multiplets are produced after the quench. 
entangled quasiparticles. However, as shown in~\cite{carollo2022entangled}, 
even for $n=3$, i.e., for entangled triplets, the tripartite information is zero. 
Let us briefly recall the proof of this result.
The only quasiparticle configuration that has to be considered is that 
with a quasiparticle in each interval $A_j$.  Without loss of generality, 
we can assume $\{a_i\}=\{1\}$, $\{b_i\}=\{2\}$, $\{c_i\}=\{3\}$ because the 
result does not depend on the permutation of the labels of the quasiparticles. 
Now, since for any system in a pure state we have that $S_A=S_{\bar A}$, 
we obtain that (see section~\ref{sec:genmethod}) $s_{\{1,2,3\}}=s_\emptyset=0$, $s_{\{1,2\}}=s_{\{3\}}$, 
$s_{\{2,3\}}=s_{\{1\}}$ and $s_{\{1,3\}}=s_{\{2\}}$. It is straightforward to 
check that this implies that $\tau_3(k)=0$ (cf.~\eqref{eq:trip_infinitesimal}). 

Thus, the simplest case in which there can be nonzero 
tripartite information is that of the entangled \textit{quadruplets} ($n=4$). 
Again, the entanglement content remains the 
same under  exchange of the quasiparticles inside and outside of the subsystem 
of interest. This implies that if all the four quasiparticles are in 
$A=A_1\cup A_2\cup A_3$, $\tau_3$ vanishes. 
Clearly, the only nontrivial configuration that contributes to $\tau_3$ is that 
with one quasiparticle in each interval $A_1,A_2,A_3$, and one quasiparticle 
outside of $A$. In the following, we are going to show that for the ``classically'' entangled 
states introduced in section~\ref{sec:positive_trip}, one has $\tau_3>0$ for any $k$, 
which implies that the tripartite information is positive at any time. Specifically, for $n=4$ 
Eq.~\eqref{eq:trip_infinitesimal} (see Fig.~\ref{fig:cartoon-2}) becomes 
\begin{multline}
	\label{eq:nonzero_quadr}
	\tau_3(k)=s_{\{1,2,3\}}(k) -s_{\{1,2\}}(k)-s_{\{2,3\}}(k)-s_{\{1,3\}}(k)\\
	+s_{\{1\}}(k)+s_{\{2\}}(k)+s_{\{3\}}(k). 
\end{multline}
Following~\cite{bertini2018entanglement}, Eq.~\eqref{eq:nonzero_quadr} 
can be rewritten (cf.~\eqref{eq:rho-in} and~\eqref{eq:e-bertini}) as 
\begin{multline}
	\label{eq:nonzero_func}
	\tau_3(k)=f(a+b+c)-f(a+b)-f(a+c)-f(b+c)\\+f(a)+f(b)+f(c),
\end{multline}
where (cf.~\eqref{eq:e-bertini}) 
	\begin{equation}
	\label{eq:quadr_abc}
	f(x)=-x \ln(x) -(1-x)\ln(1-x),
\end{equation}
with 
\begin{equation}
	\label{eq:abc-quartet}
	a=\rho_1(k),\quad b= \rho_2(k),\quad c= \rho_3(k).
\end{equation}
The total density is constrained as $\sum_{j=1}^n\rho_j(k)=1$, 
and the variables $a, b, c$ satisfy $a\geq0$, $b\geq0$, $c\geq0$ and 
$a+b+c\leq1$. 
The expression in~\eqref{eq:nonzero_func} is always non-negative 
under the given constraints on the densities $a$, $b$ and $c$. 
Indeed, one can easily check that  $\tau_3$ (cf.~\eqref{eq:nonzero_func}) 
is smooth as a function of $a,b,c$, and it vanishes at the boundaries of the allowed region 
for $a,b,c$. Moreover, Eq.~\eqref{eq:nonzero_func}  has 
a unique stationary point at $a=b=c=1/4$, where it is positive. 
This allows us to conclude that  $\tau_3>0$ for any $a,b,c$, except at the 
boundaries where $\tau_3=0$. 
Notice that the boundaries ($a=0$, $b=0$, $c=0$)  correspond to the cases 
with at least one of the intervals $A_1,A_2,A_3$ not containing a quasiparticle, 
that do not contribute to the TMI. 

Let us now consider  the general case with arbitrary $n$-plets with $n>4$. 
Specifically, let us consider the situation in which  quasiparticles with 
indices $\{a_j\}_{j=1}^p$ are in $A_1$, those with $\{b_j\}_{j=1}^q$ in $A_2$, 
and with $\{c_j\}_{j=1}^r$  in $A_3$. We have $p+q+r\le n$. 
Crucially, Eq.~\eqref{eq:trip_infinitesimal} has the same form 
as~\eqref{eq:nonzero_func} with different $a$, $b$ and $c$, that are defined as  
\begin{equation}
	\label{eq:gen_abc}
	a=\sum_{j=1}^p\rho_{a_j}(k),\;\;
	b= \sum_{j=1}^p\rho_{b_j}(k),\;\;
	c= \sum_{j=1}^r\rho_{c_j}(k).
\end{equation}
Moreover, the $a$, $b$, $c$ in~\eqref{eq:gen_abc} 
satisfy the same constraint, i.e., $a\geq0$, $b\geq0$, $c\geq0$, $a+b+c\leq1$,  
as in the case of quadruplets (cf.~\eqref{eq:abc-quartet}). 
This implies that $\tau_3(k)\ge 0$ for any $k$, which allows us to 
conclude that the TMI cannot be negative for the ``classically'' entangled 
states of Ref.~\cite{bertini2018entanglement}.

%####################################################
\section{Negative TMI after a quench in the $XX$  chain}
\label{sec:negative_free}
	
Having established in the previous section that quenches starting from states of the form~\eqref{eq:free_states} in the free fermion 
chain studied in Ref.~\cite{bertini2018entanglement}  
give rise to a non-negative tripartite mutual information, 
we now provide a setup in which  $I_3(t)$ is \textit{negative} at intermediate times 
in the hydrodynamic limit. 

Precisely, let us now consider the quench in the $XX$ chain starting from 
the state $|\Phi_0\rangle$ (cf.~\eqref{eq:ex_init}). 
The state exhibits a four-site translation invariance. The dynamics from $|\Phi_0\rangle$ 
produces entangled quadruplets, and in contrast with the states considered in Ref.~\cite{bertini2018entanglement}, 
contains two fermions per unit cell. This implies that the 
correlation matrix $G_{ij}(k)$ (cf.~\eqref{eq:Gij}) has rank larger than one 
(the last step in~\eqref{eq:Gmat_rk1_calc} does not hold). 
Crucially, this means that the von Neumann entropy is not straightforwardly obtained from 
the fermionic occupations $\rho_j(k)$ (cf.~\eqref{eq:rhoj}), i.e., from the GGE that describes 
the steady state. 

Before proceeding, let us observe that since the initial state has a well defined fermion number, one has 
that $F_{ij}(k)=0$ (cf.~\eqref{eq:Fij}) at any time after the quench. 
Now, we restrict the Brillouin zone to ${\mathcal B}=(\pi/2,\pi]$, and define the 
four quasiparticles $\eta_j(k)$, $j\in[1,4]$ according to~\eqref{eq:freechain_redef}. 
The associated group velocities are 
\begin{equation}
	\label{eq:freeneg_v}
		v_j(k)=\frac{d}{dk}\varepsilon_j(k)=2 \sin\left(k-(j-1)\frac{\pi}{2}\right), 
	\end{equation}
where $\epsilon_j(k)$ are the dispersions of the different 
species (cf.~\eqref{eq:freechain_redef}). As it is clear from~\eqref{eq:freeneg_v},  
$v_1$ and $v_2$ are positive in the reduced Brillouin zone, while $v_3=-v_1$ and $v_4=-v_2$. 

Furthermore, a straightforward calculation gives the fermionic correlation matrix 
$G(k)$ (cf.~\eqref{eq:Gij}) as 
\begin{equation}
		\label{eq:freeneg_G}
		G(k)=\frac{1}{4}\begin{pmatrix}
		2&-1-i&0&-1+i\\
		-1+i&2&-1-i&0\\
		0&-1+i&2&-1-i\\
		-1-i&0&-1+i&2\\
		\end{pmatrix}.
\end{equation} 
Notice that $G(k)$ does not depend on $k$, similarly to the quench from 
the fermionic N\'eel state~\cite{mazza2016overlap}. 

%
%####################################################
\begin{figure}[t]
\includegraphics[width=.48\textwidth]{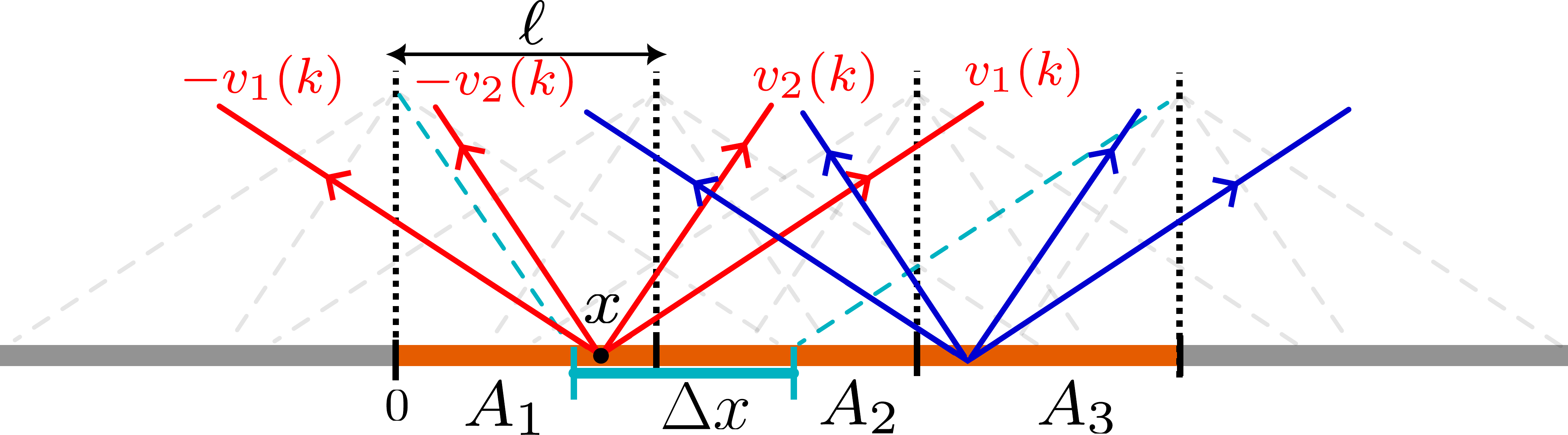}
\caption{A typical entangled multiplet with $\nu=4$ contributing 
 to the dynamics of the tripartite mutual information TMI in the 
 $XX$ chain after the quench from the state 
 $|\Phi_0\rangle=\left|\uparrow\uparrow\downarrow\downarrow\right
 \rangle^{\otimes L/4}$. 
 We consider the TMI $I_3$ between three equal intervals $A_i$ of 
 length $\ell$. We show the contribution of an entangled quadruplet 
 produced at a generic position $x$. Here we consider the case with 
 the group velocities of the quasiparticles being 
 $v_1(k),v_2(k)\ge 0$, $v_3(k)=-v_1(k)$ and 
 $v_4(k)=-v_2(k)$. The different colors show the two types of 
 quadruplets that contribute to $I_3$. They correspond to the 
 situation in which $A$ is entangled with $\bar A$ via the left 
 and right boundary, respectively. For the first case, the 
 total number of multiplets contributing to $I_3$ is proportional to the 
 width $\Delta x$. 
}
\label{fig:cartoon-3}
\end{figure}
%####################################################
%

We are now ready to discuss the quasiparticle picture for the dynamics of the TMI.  
The direct calculation of the tripartite information 
within the quasiparticle picture is somewhat easier than the 
calculation of the entropies. 
Specifically, the reason is that the only quasiparticle configurations 
yielding nonzero TMI are those with exactly one particle inside each of the three intervals 
and one outside of $A$. From the velocities~\eqref{eq:freeneg_v} 
it is straightforward to realize that there are only four ways 
to satisfy this condition, which  
depend on the ordering of the velocities. Specifically, we have to 
consider the two cases as 
\begin{itemize}
	\item[$(i)$] for $\pi/2\le k\le 3/4\pi$, one has 
	$v_1(k)\ge v_2(k)\ge 0$ and $v_3(k)\le v_4(k) \le 0$.  
	Now, there are only two possibilities to have nonzero $I_3$. 
	The first one is  that the quasiparticle of species $1$ is in  $A_3$,  
	that of species $2$ is in  $A_2$, and that of species  
	$4$ in $A_1$, with the quasiparticle of species 
	$3$ outside of $A$ on the left. 
	The other possibility is that the quasiparticle of species $1$ is outside 
	of $A$ on the right, that of species  $2$ is in $A_3$, that of species 
	$4$ in $A_2$, and that of species  $3$ is in $A_3$. These two 
	configurations are depicted in Fig.~\ref{fig:cartoon-3} with different 
	colors. 
\item[$(ii)$] for $3/4\pi\le k\le \pi$ one has 
$v_2(k)\ge v_1(k)>0$ and $v_4(k)\le v_3(k)\le 0$. 
The configurations that contribute to the TMI are the same as in $(i)$ after 
the exchanges $1\leftrightarrow 2$ and $3\leftrightarrow 4$. 
\end{itemize}
%
%
%####################################################
\begin{figure}
\centering
\includegraphics[width=1.05 \linewidth]{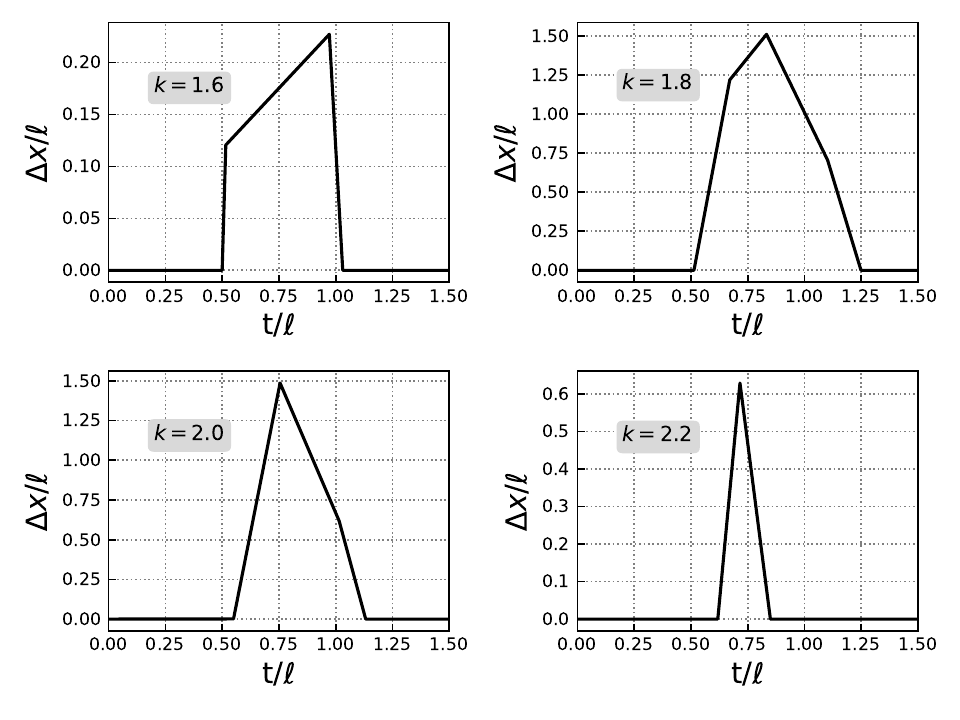}
\caption{ Tripartite mutual information (TMI) in the $XX$ chain after 
 the quench from the state $|\Phi_0\rangle$ (cf.~\eqref{eq:ex_init}). We show 
 $\Delta x/\ell$, with $\Delta x={\mathcal D}_1+{\mathcal D}_2$ (cf.~\eqref{eq:freeneg_trip}). 
 We plot $\Delta x/\ell$ for fixed quasimomentum $k$ versus $t/\ell$. 
 Notice that apart for a constant, $\Delta x$ is the contribution of the quasiparticles 
 to $I_3$. The different panels correspond to different $k$. The cusp-like features 
 are due to the presence of quasiparticles with different velocities. 
}
\label{fig:xx-cinem}
\end{figure}
%####################################################
%

It is straightforward to obtain the total number of configurations 
contributing to $I_3$. Let us focus on the first  
type $(i)$. Let us consider an entangled multiplet produced at a generic 
point $x$ at $t=0$. We only consider the situation in which the leftmost 
quasiparticle is outside of $A$ on the left (see Fig.~\ref{fig:cartoon-3}). 
The conditions that give nonzero $I_3$ are
\begin{equation}
 x\ge v_2t,\quad x\ge \ell -v_2t,\quad  x\ge 2\ell-v_1t, 
\end{equation}
together with 
\begin{equation}
	x\le v_1t,\quad x\le\ell+v_2t,\quad x\le 2\ell-v_2t,\quad x\le 3\ell-v_1t. 
\end{equation}
As it is clear from Fig.~\ref{fig:cartoon-3}, the constraints above identify the region 
of width $\Delta x$ in which the quadruplets that contribute to 
$I_3$ are produced at $t=0$. 
After integrating over all the possible positions $x$, one obtains 
\begin{multline}
	\label{eq:d1}
	{\mathcal D}_1(k,\ell,t)=
	\max\{\min \{v_1t, \ell+v_2t,2\ell-v_2t, 3\ell-v_1t\}\\-
	\max \{ v_2t,\ell-v_2t, 2\ell-v_1t\},0\},
\end{multline}
The entangled quadruplets of type $(i)$ in which the 
rightmost particle is outside of $A$ give 
\begin{multline}
	\label{eq:d2}
	{\mathcal D}_2(k,\ell,t)=
	\max\{\min \{\ell+v_1t, 2\ell+v_2t,3\ell-v_2t\}\\-
	\max \{ v_1t,\ell+v_2t, 2\ell-v_2t,3\ell-v_1t\},0\}. 
\end{multline}
Together with~\eqref{eq:d1} and~\eqref{eq:d2}, there are two contributions 
${\mathcal D}_3$ and ${\mathcal D}_4$, which are obtained by exchanging 
$v_1\leftrightarrow v_2$ and $v_3\leftrightarrow v_4$. 

To proceed, 
we now determine the contribution of the entangled multiplets to the 
TMI. This is straightforward using the strategy discussed in section~\ref{sec:genmethod}. 
Specifically, by using~\eqref{eq:s_A} and~\eqref{eq:freeneg_G} one can verify  that the 
contribution $\tau_3(k)$ (cf.~\eqref{eq:trip_infinitesimal}) does not depend on $k$ 
and on the different ways ${\mathcal Q}_A$ of distributing the quasiparticles in 
the subsystems. We obtain $\tau_3$ as 
\begin{equation}
\label{eq:tau-3-xx}
\tau_3=2 f\left(\frac{1}{2}\right)-4 
f\left(\frac{2+\sqrt{2}}{4}\right)<0, 
\end{equation}
where $f(x)$ is given in~\eqref{eq:quadr_abc}. 
Crucially, $\tau_3$ is negative for any $k$.  
Putting together~\eqref{eq:d1}~\eqref{eq:d2} and~\eqref{eq:tau-3-xx}, we 
obtain 
\begin{multline}
\label{eq:freeneg_trip-1}
I_3(t)=\Big[ 2 f\Big(\frac{1}{2}\Big)-4 f\Big(\frac{2+\sqrt{2}}{4}\Big)\Big]\\
\times \Big( \int_{\pi/2}^{3\pi/4}\frac{dk}{2\pi} [\mathcal{D}_1(k,\ell,t)+
\mathcal{D}_2(k,\ell,t)] 
\\+\int_{3\pi/4}^{\pi}\frac{dk}{2\pi} [\mathcal{D}_3(k,t,\ell)+\mathcal{D}_4(k,t,\ell)]\Big). 
\end{multline}
Finally, the two terms in~\eqref{eq:freeneg_trip-1} give the same result. Thus, 
we can rewrite~\eqref{eq:freeneg_trip-1} as 
\begin{multline}
\label{eq:freeneg_trip}
I_3(t)=\Big[ 4 f\Big(\frac{1}{2}\Big)-8 f\Big(\frac{2+\sqrt{2}}{4}\Big)\Big]\\
\times \int_{\pi/2}^{3\pi/4}\frac{dk}{2\pi} [\mathcal{D}_1(k,\ell,t)+
\mathcal{D}_2(k,\ell,t)] 
\end{multline}
We should mention that the terms in $\tau_3(k)$ in~\eqref{eq:tau-3-xx} appear naturally 
in the quasiparticle picture for the von Neumann entropy of a single interval (see Appendix~\ref{sec:app} for 
an \emph{ab initio} derivation). 
Let us discuss the dynamics of $I_3$ as obtained from~\eqref{eq:freeneg_trip}. 
In the following we refer to times such that $t,\ell\to\infty$ with $t/\ell\ll 1$ as short times, whereas 
by asymptotically long times we mean the situation with $t,\ell\to\infty$ with $t/\ell\to\infty$. 
At short times,  
$I_3=0$, and it remains zero up to time 
$t=\ell/(\max(v_1(k),v_2(k)))$, when the quasiparticles forming the quadruplets start 
being shared between all the subsystems. At later times, $I_3$ decreases, reaching a 
minimum. Finally, it vanishes at asymptotically long times, when the particles of each multiplet 
are too far from each other to be shared between all the subsystems. It is interesting to 
investigate the behavior of the integrand in~\eqref{eq:freeneg_trip} as a 
function of time. As it is clear from the derivation of~\eqref{eq:d1} and~\eqref{eq:d2}, 
the integrand is the width of the spatial region where the entangled multiplets that 
at a given time give nonzero $I_3$ are produced. 
In Fig.~\ref{fig:xx-cinem} we report $\Delta x/\ell$, with $\Delta x={\mathcal D}_1+{\mathcal D}_2$. 
As anticipated, 
$I_3$ is zero at short times. This corresponds to the fact that at short times there 
are no entangled quadruplets that are shared among the three subsystems $A_j$ and 
the rest. Moreover, one should observe that at intermediate times the behavior 
of $I_3$ is quite involved and it depends dramatically on the quasimomentum $k$. 
Specifically, $\Delta x/\ell$ exhibits several cusp-like features. These reflect the 
fact that different quasiparticles in the same multiplet have different velocities. 
Notice that these cusp-like features could be detected in numerical studies by 
monitoring the behavior of $d I_3/dt$, similar to what observed for the 
von Neumann entropy~\cite{fagotti2008evolution}.

%####################################################
\section{Ising chain with  staggered transverse magnetic field}
\label{sec:ising}

%####################################################
\begin{figure}[t]
\includegraphics[width=.5\textwidth]{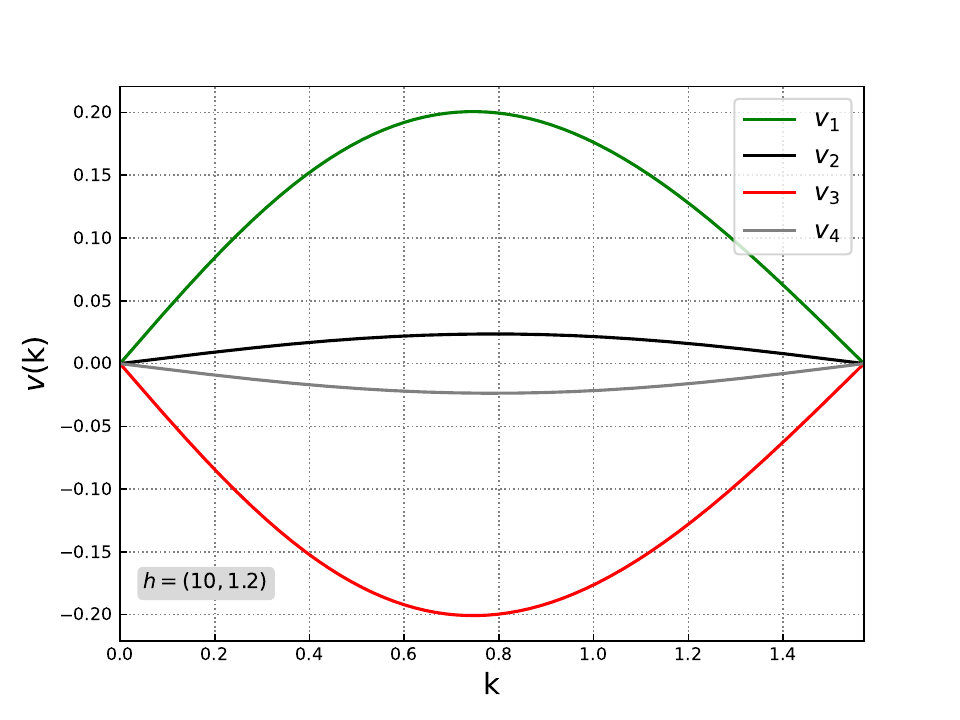}
\caption{ Group velocities $v_j(k)$ of the quasiparticles forming entangled quadruplets in the 
 Ising chain with $h=(10,1.2)$. We plot $v_j(k)$ versus the 
 quasimomentum $0\le k\le \pi/2$. Notice that  $v_1(k)>v_2(k)>v_4(k)>v_3(k)$ 
 for any $k$ except at $k=0,\pi/2$, where they all vanish. 
}
\label{fig:isingvk}
\end{figure}
%####################################################

Here we derive the quasiparticle picture for $I_3$  after a quench in the transverse 
field Ising chain with staggered magnetic field (cf.~\eqref{eq:ham-is} 
and~\eqref{eq:ising-c}). We restrict ourselves to the situation in 
which the magnetic field has periodicity two, with values $h=(h_o,h_e)$, 
where $h_o$ and $h_e$ is the magnetic field on the odd and even sites of 
the chain, respectively. We consider the following quench protocol. At $t=0$ 
the chain is prepared in the ground state of the model with $h^0=(h_o^0,h_e^0)$. 
At $t>0$, the magnetic field is suddenly changed to $h=(h_o,h_e)$, and the 
system evolves with the new Hamiltonian. In the following we show that, similarly to 
the $XX$ quench discussed in section~\ref{sec:negative_free}, this will give rise to 
a negative TMI. 

To find an explicit form for the two-body correlation function $\mathcal{C}(k)$ in~\eqref{eq:2pt_Cblock} 
for generic magnetic fields $h$, we have to diagonalize the $4\times4$ matrix $\mathcal{H}_k$
in~\eqref{eq:h-supp} for both $h$ and $h^0$, thus determining ~\eqref{eq:2pt_Cblock} 
via equation~\eqref{eq:unitdiag}. Although it is, in principle, possible to 
analytically perform the diagonalization in our specific case of a two-site periodic field, 
the expressions for the eigenvalues $\varepsilon_i(k)$ and the eigenvectors as functions 
of $h$, $h^0$ and $k$ are very cumbersome and not particularly enlightening.
Thus, we prefer to perform the diaognalization numerically. From the eigenvalues $\varepsilon_i(k)$, one obtains the 
group velocities of the quasiparticles as $v_i(k)=d \varepsilon_i(k)/dk$. For the following, it is useful to observe that $v_i(k)=-v_{i+n}(k)$, 
because the eigenvalues of $\mathcal{H}_k$ are organized in pairs with opposite signs (see~\eqref{eq:ising_diag}). 
In Fig.~\ref{fig:isingvk} we report the group velocities $v_j(k)$ for the Ising chain with $n=2$ and $h=(10,1.2)$. The quasimomentum $k$ of 
the species is in $[0,\pi/2]$. Notice that the order of the velocities associated to the various quasiparticle species is the same for all 
the quasimomenta. The same holds for all the values of $h$ we take into account in the following.
This means that  the kinematics of the quasiparticles is qualitatively the same as in the $XX$ chain after the 
quench discussed in section~\ref{sec:negative_free}, and we have the same scenario as in Fig.~\ref{fig:cartoon-3}. 
In the hydrodynamic limit, $I_3$ is thus described by the formula 
\begin{equation}
\label{eq:ising_trip}
I_{3}(t)=
\int_{0}^{\pi/2}\frac{dk}{2\pi} \tau_3(k) (\mathcal{D}_1(k,\ell,t)+\mathcal{D}_2(k,\ell,t)),
\end{equation}
where the functions $\mathcal{D}_1$ and $\mathcal{D}_2$ are the same as 
in~\eqref{eq:freeneg_trip} if we label the quasiparticle species  so that $v_1>v_2$.
Here the entanglement content $\tau_3(k)$ is the same as in~\eqref{eq:trip_infinitesimal}, 
where $s_{\{x\}}(k)$ is the entropy obtained numerically from 
${\mathcal C}(k)$ (cf.~\eqref{eq:2pt_Cblock}) as explained in section~\ref{sec:genmethod}. 
Clearly, now $\tau_3$ depends on $k$, in contrast with the 
case of the $XX$ chain (see section~\ref{sec:negative_free}). 

\begin{figure}
\centering
\includegraphics[width=1 \linewidth]{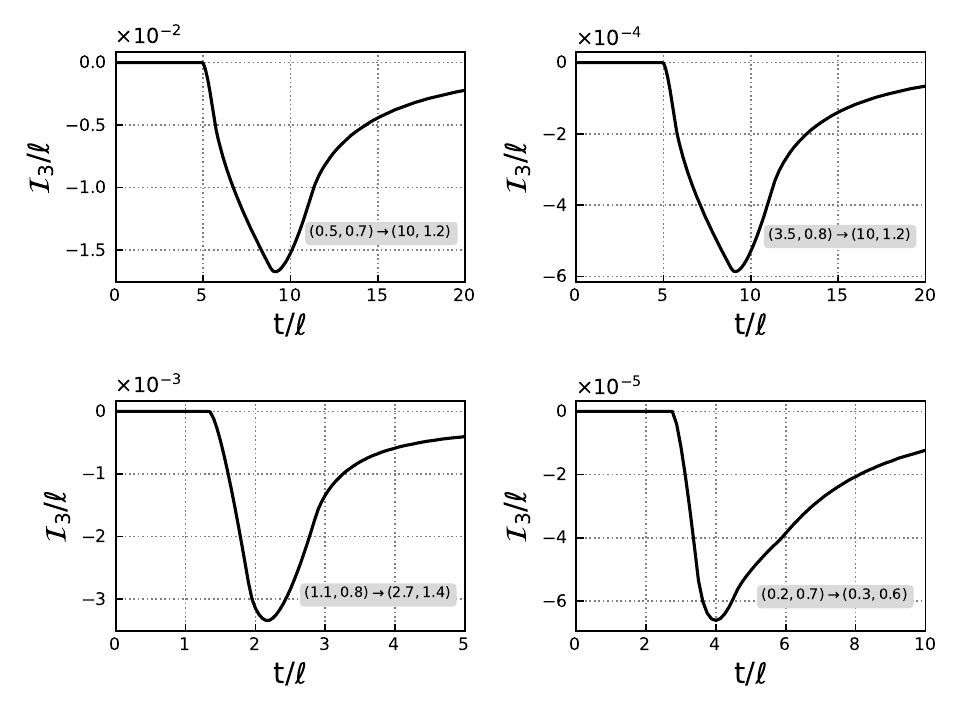}
\caption{ Tripartite mutual information (TMI) between  
 three adjacent intervals of length $\ell$ after a quench in the transverse field 
 Ising chain. We plot the density $I_3/\ell$ for TMI versus the rescaled time $t/\ell$. 
 The different panels correspond to different quenches $(h_o^0,h_e^0)\to(h_o,h_e)$. Notice 
 that the $y$-axis is  rescaled, the rescaling factor being reported at the 
 top of each panel. 
}
\label{fig:ising_pred}
\end{figure}

In Figure~\ref{fig:ising_pred}, we show the quasiparticle prediction for $I_3$ 
in the Ising chain after several quenches $h^{(0)}\to h$ (different panels in the figure). 
The results are for three adjacent intervals of equal length $\ell$. Again, the quasiparticle 
picture holds in the hydrodynamic limit $\ell,t\to\infty$ with the ratio $t/\ell$ fixed. 
Interestingly, for all the quenches that we analyzed the TMI attains quite ``small'' values 
$\lesssim 10^{-2}$ . 
The TMI is zero at short times. Precisely, one has $I_3/\ell=0$ for $t/\ell\le 1/v_\mathrm{max}$. 
As it is clear from Fig.~\ref{fig:isingvk} one has $v_\mathrm{max}\approx 0.2$ for the quench 
with $h=(10,1.2)$, which implies that $I_3=0$ for $t/\ell\lesssim 5$. 
At $t/\ell=1/v_\mathrm{max}$ the TMI starts decreasing. This happens because an 
entangled quadruplet created at the boundary between $A_1$ and $A_2$ (or $A_2$ and $A_3$) 
starts to be shared, and hence it contributes to $I_3$. 
Quite generically, at later times $I_3$ is negative, and becomes smaller and smaller upon increasing times. 
Thus, $I_3$ reaches a minimum, and then starts growing. At asymptotically long times 
$I_3$ vanishes. The vanishing of the TMI signals that, although the multiplets exhibit 
nontrivial correlations, the dynamics of the quantum information shared between the 
different intervals happens in a ``localized'' manner via the propagation of the quasiparticles. 
The vanishing of $I_3$ reflects that at infinite times there are no entangled quadruplets that 
are shared between $A_1,A_2,A_3$ and the complement of $A$. 

%
%####################################################
%####################################################
%

%####################################################
\section{Numerical benchmarks}
\label{sec:num}
	
In this section we benchmark our predictions~\eqref{eq:freeneg_trip} 
and~\eqref{eq:ising_trip} against numerical simulations. 
Again, we focus on the hydrodynamic limit. In section~\ref{sec:xx_num} we discuss 
the case of the $XX$ chain, whereas in section~\ref{sec:ising_checks} we 
consider the Ising chain. In both sections we consider the situation with 
three  adjacent intervals $A_1,A_2,A_3$ of equal length $\ell$. We discuss data 
for $\ell\lesssim 400$. Our numerical results for the TMI in the $XX$ chain 
are obtained by using~\eqref{eq:2pt_Cblock} where $F_{ij}=0$ and $G_{ij}$ is given 
in~\eqref{eq:freeneg_G}. 
The starting point is the real space correlator ${\mathcal C}$ for the full chain,
which satisfies a system of $L^2$ linear differential equations. 
These equations can be efficiently solved, for instance by Fourier transform, 
allowing one to obtain the dynamics of the correlator. 
From that, one can obtain the von Neumann entropy of any subsystem at any time by using the method of Ref.~\cite{peschel2009reduced}. 
The calculation of the TMI is then straightforward by using~\eqref{eq:trip-def}. 
A similar, although slightly more involved, procedure can be employed for the Ising chain.

%####################################################
\subsection{XX chain}
\label{sec:xx_num}

%
%####################################################
\begin{figure}
\centering
\includegraphics[width=.5 \textwidth]{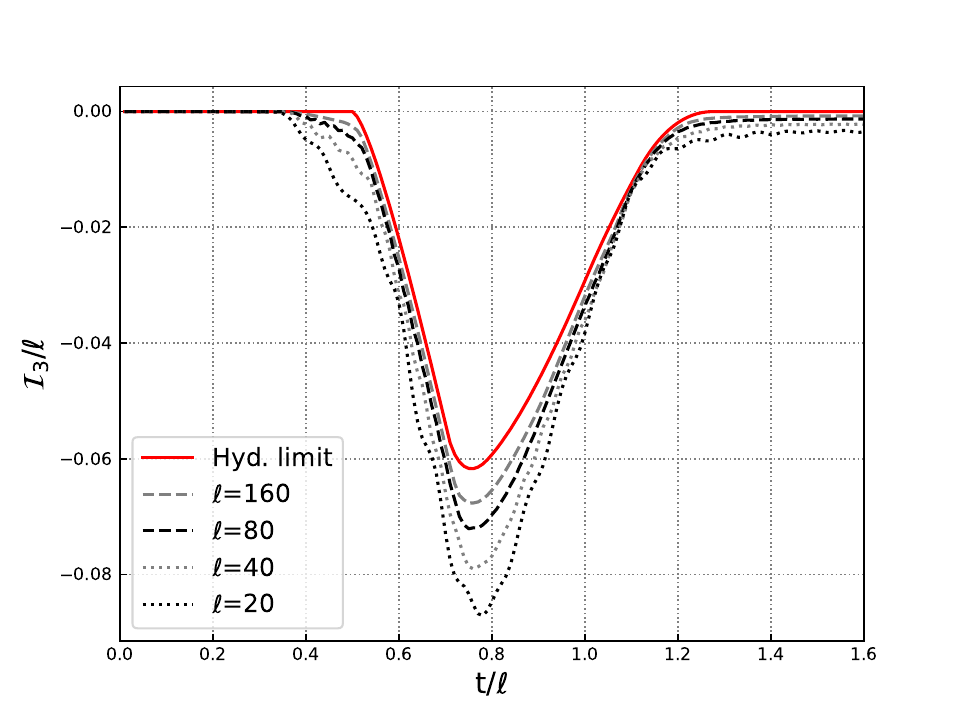}
\caption{Dynamics of the tripartite mutual information $I_3$ in the $XX$ chain 
 after the quench from the state $|\Phi_0\rangle$ (cf.~\eqref{eq:ex_init}). 
 We show results in the hydrodynamic limit. 
 The different lines are for different lengths $\ell$ of the intervals. 
 The continuous line (red line) is the prediction of the quasiparticle 
 picture~\eqref{eq:freeneg_trip}.
}
\label{fig:freeneg_conv}
\end{figure}
%####################################################
%
%####################################################
\begin{figure}
\centering
\includegraphics[width=1 \linewidth]{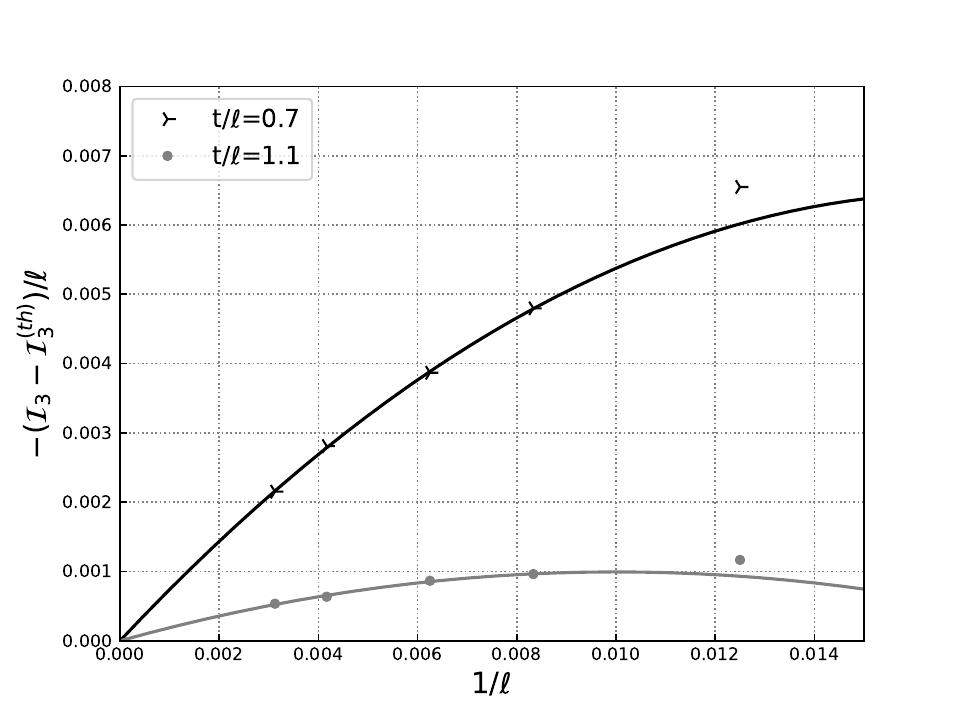}
\caption{Dynamics of $I_3$ after the quench from the state $|\Phi_0\rangle$ in 
 $XX$ chain. Finite-size corrections to the hydrodynamic limit. 
 The figure shows the difference $(I_3-I_3^{\scriptscriptstyle(th)})/\ell$, 
 with $I_3^{\scriptscriptstyle(th)}$ being the quasiparticle prediction for $I_3$ in 
 the hydrodynamic limit. The $x$-axis shows $1/\ell$, with $\ell$ being the 
 length of the intervals. The symbols in the figure are the results at 
 fixed $t/\ell$. The full lines are fits to $a/\ell+b/\ell^2$, with $a,b$ fitting 
 parameters. For both values of $t/\ell$ the rightmost point is 
 excluded from the fit. 
}
\label{fig:freeneg_singlepoint}
\end{figure}
%####################################################
%

Our numerical results for the $XX$ chain are reported in Fig.~\ref{fig:freeneg_conv}. 
The figure shows numerical data for $I_3/\ell$ plotted as a function of $t/\ell$. 
The initial state of the quench is $|\Phi_0\rangle$ (cf.~\eqref{eq:ex_init}). 
Fig.~\ref{fig:freeneg_conv} shows that even for finite $\ell$ the TMI is negative at 
all times. However, deviations from the quasiparticle picture (cf.~\eqref{eq:freeneg_trip} and 
continuous red line in Fig.~\ref{fig:freeneg_conv}) are 
visible. Upon increasing $\ell$ the numerical data approach the analytic result. 
A more systematic analysis is reported in Fig.~\ref{fig:freeneg_singlepoint} where 
we show data for $(I^{\scriptscriptstyle(th)}_3-I_3)/\ell$ at fixed $t/\ell=0.7$ and $t/\ell=1.1$ plotted versus $1/\ell$. 
We show data for $\ell\lesssim 400$. 
Here $I_3^{\scriptscriptstyle (3)}$ is~\eqref{eq:freeneg_trip}. The continuous lines in 
Fig.~\ref{fig:freeneg_singlepoint} are fits to $a/\ell+b/\ell^2$, with $a,b$ fitting 
parameters. The functional form of the fitting function is motivated by the fact that 
similar corrections are observed for the von Neumann entropy~\cite{fagotti2008evolution}. 
Moreover, such corrections appear naturally in the stationary phase approximation~\cite{wong} that 
is used to derive~\eqref{eq:freeneg_trip}.

%####################################################
\subsection{Ising chain}
\label{sec:ising_checks}
	
Let us now discuss the behavior of $I_3$ after a quench in the Ising chain with 
staggered magnetic field (see section~\ref{sec:models}). Here we consider the 
case with $h^{(0)}$ and $h$ taking different values on the odd and even sites of the 
lattice. The quench protocol is as follows. The chain is initially prepared in the 
ground state of the Ising chain with $h^{(0)}$. At $t=0$ the magnetic field is suddenly 
quenched to the final value $h$, and the system evolves with the new Hamiltonian.

\begin{figure}
\centering
\includegraphics[width=1 \linewidth]{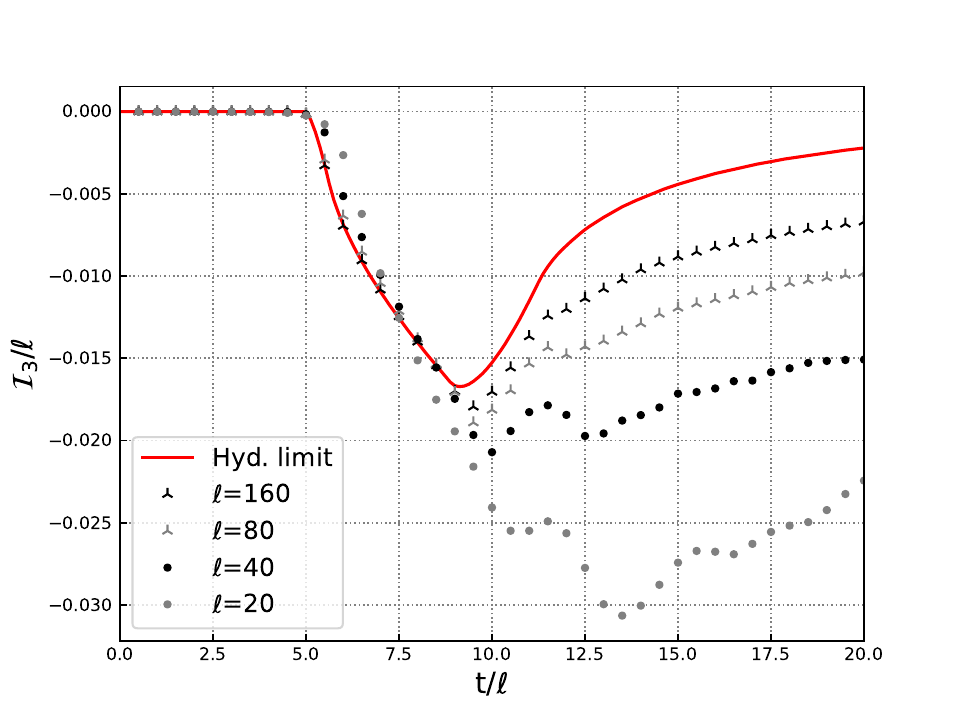}
\caption{ Dynamics of $I_3$ after a quench in the Ising chain with 
 staggered transverse field. Here we consider the quench $h^{(0)}=(0.5,0.7)\to h=(10,1.2)$. 
 The figure shows $I_3/\ell$ for the geometry with three adjacent intervals of equal size 
 $\ell$ (see Fig.~\ref{fig:cartoon}). The continuous red line is the prediction in the 
 hydrodynamic limit, equation~\eqref{eq:ising_trip}. 
 Notice that at finite $\ell$ the data exhibits strong oscillating corrections. 
}
\label{fig:ising_conv}
\end{figure}

\begin{figure}
\centering
\includegraphics[width=1 \linewidth]{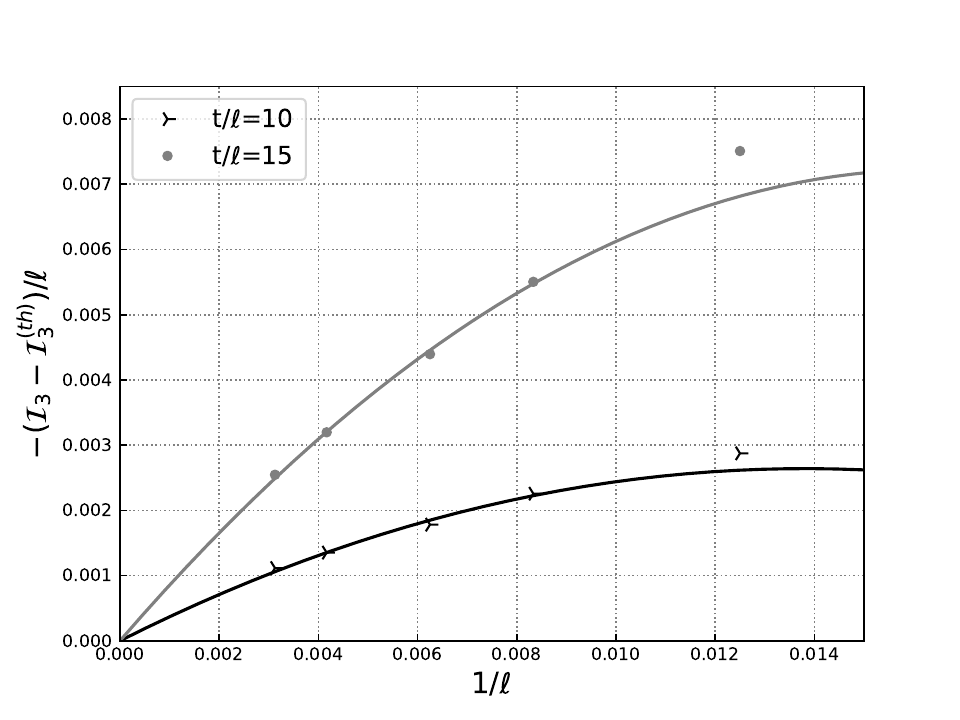}
\caption{ Dynamics of $I_3$ after a magnetic field quench in the transverse field 
 Ising chain. The setup is the same as in Fig.~\ref{fig:ising_conv}. We show the finite-size 
 corrections to $I_3$ plotting $I_3^{\scriptscriptstyle(th)}-I_3$, with $I_3^{\scriptscriptstyle(th)}$ 
 given by~\eqref{eq:ising_trip}, versus $1/\ell$ at fixed $t/\ell=10,15$. The continuous lines are 
 fits to $q/\ell+b/\ell^2$, with $a,b$ fitting parameters. For both values of $t/\ell$ the rightmost point is 
 excluded from the fit. 
}
\label{fig:ising_singlepoint}
\end{figure}

In Fig.~\ref{fig:ising_conv} we show numerical results for $I_3$ for the quench 
$(0.5,0.7)\to (10,1.2)$. Now, the finite-size data exhibit sizable deviations 
from the analytic result in the hydrodynamic limit (reported as continuous red curve 
in Fig.~\ref{fig:ising_conv}). Moreover, the data show a clear oscillating behavior 
as a function of time. Still, upon increasing $\ell$ the numerical results approach 
the analytic curve. In Fig.~\ref{fig:ising_singlepoint} we perform a finite-size 
scaling analysis plotting $I_3^{\scriptscriptstyle(th)}-I_3$, where $I_3^{\scriptscriptstyle(th)}$ 
is the hydrodynamic formula~\eqref{eq:ising_trip}. As for the $XX$ chain (see Fig.~\ref{fig:freeneg_singlepoint}), 
the continuous lines are fits to $a/\ell+b/\ell^2$. The quality of the fits is satisfactory, 
confirming the validity of~\eqref{eq:ising_trip}. 

%
%####################################################
%
%####################################################
%
%####################################################
%####################################################
%

%####################################################
\section{Conclusions}
\label{sec:concl}

We derived a quasiparticle picture description for the dynamics of 
the tripartite information after quantum quenches in the $XX$ chain and 
the Ising chain with staggered magnetic field. Precisely, we focused on the 
situation in which entangled multiplets are produced after the quench. 
In the presence of entangled pairs (or triplets) of quasiparticles, the TMI vanishes in 
the hydrodynamic limit of long times and large subsystems, with their ratio fixed. 
Instead, if entangled multiplets with  more than three particles are present, 
the TMI is nonzero. Moreover, for the  entangled multiplets 
investigated in Ref.~\cite{bertini2018entanglement} the TMI is positive at 
intermediate times. This reflects that the quasiparticles forming the entangled 
multiplets are only ``classically'' correlated, and the dynamics of the 
von Neumann entropy and of the TMI are describable in terms of GGE thermodynamic 
information~\cite{bertini2018entanglement}. In contrast, we showed that 
if the quasiparticles forming the entangled multiplets are nontrivially correlated, 
the TMI is negative at intermediate times. In the latter case, a hydrodynamic description 
of the TMI is still possible. However, the correlation content of the multiplets 
is not given in terms of the GGE, although it can be determined with modest 
computational cost for systems that are mappable to free fermions.
The relationship between the entanglement content and the GGE 
is recovered only in the limit of long times, $t/\ell \to \infty$, when the distance 
between the quasiparticles forming a multiplet is large, and only one quasiparticle 
can be in the subsystem. 

Our work opens several interesting research avenues. First, it is important 
to further investigate the relationship between the sign of the TMI and the 
structure of the entangled multiplets. Specifically, it would be interesting 
to understand under which conditions on the fermionic correlation 
matrix~\eqref{eq:2pt_Cblock} the TMI is negative. 
While we showed that genuine quantum correlation between the 
quasiparticles forming the multiplets is necessary to have negative TMI, it is not 
clear whether the converse is true. It would be interesting to understand whether 
it is possible to have nontrivially entangled multiplets giving a positive TMI. 
Also, it would be interesting 
to investigate the behavior of the TMI in free-boson systems~\cite{alba2018entanglement}. 
An important direction would be to extend our results to free-fermion and free-boson 
systems in the presence of dissipation. 
It has been shown~\cite{alba2021spreading,carollo2022dissipative,alba2022hydrodynamics,alba2022logarithmic,murciano2023symmetryresolved} 
that for quadratic Markovian dissipative dynamics it is possible to employ the 
quasiparticle picture to describe the dynamics of entanglement-related 
quantities. Unfortunately, so far only quenches giving rise to entangled pairs were 
explored. It would be interesting to understand whether the dissipative quasiparticle 
picture can be generalized to the case of 
entangled multiplets. A crucial question is how dissipative processes affect the 
TMI. Furthermore, it is of paramount importance to understand the effect of interactions, 
although this is a formidable task. A possibility is to study the dynamics from~\eqref{eq:ex_init} in 
the $XXZ$ spin chain, which is interacting. However, it is not clear that the 
dynamics ensuing from~\eqref{eq:ex_init} can be described in terms of multiplets. Moreover, 
to build a quasiparticle picture for the TMI, or even for the entanglement entropy, one has 
to determine the correlations between the quasiparticles forming the multiplet, which is a 
nontrivial task. 
Finally, it would be interesting to understand to which extent it is possible to 
recover the quasiparticle picture from the ballistic fluctuation theory~\cite{delvecchio2023entanglement} 
in the presence of entangled multiplets.

\begin{acknowledgments}
	We would like to thank Paola Ruggiero for bringing to our attention Ref.~\cite{bastianello2018spreading} and 
	for discussions. We would also like to thank Federico Carollo for discussions in a related 
	project. 
\end{acknowledgments}

\appendix

%####################################################
\section{Single-interval entropy in the presence of quadruplets: An ab-initio derivation}
\label{sec:app}

In the following we provide an \emph{ab initio} derivation 
of the quasiparticle picture for the von Neumann entropy of 
an interval $A$ embedded in an infinite system after the quench 
from $|\Phi_0\rangle$ (cf.~\eqref{eq:ex_init})in the $XX$ chain. 

Specifically, we consider the von Neumann entropy of an interval 
of size $\ell$ at a time $t$ after the quench. We consider the 
hydrodynamic limit. 

The computation of the entropy of a subsystem $A$ for a Gaussian 
fermionic state with well-defined  particle number relies on the well-known formula 
\begin{equation}
\label{eq:ent_2pt}
S_{A}=-\mathrm{Tr}[G_{A}\ln(G_{A})+ 
	(\mathbb{1}-G_{A}) \ln (\mathbb{1}-G_{A})],
	\end{equation}
where the matrix $G$ is the two-point fermionic correlation function in the real 
space
\begin{equation}
	\label{eq:Gmat}
		G_{x,y}(t)=\mathrm{Tr}[c_x^\dagger c_y \rho(t)]=\langle \Phi_0|e^{iHt}c_x^\dagger c_y e^{-iHt}|\Phi_0\rangle,
\end{equation}
and the subscript ${A}$ is to stress that we restrict to 
subsystem ${A}$. To obtain the hydrodynamic limit 
of~\eqref{eq:ent_2pt}, we start from the moments $\mathrm{Tr}[G_{A}^n]$. 
By knowing the analytic dependence on $n$ of the moments, it is possible to 
obtain  the hydrodynamic limit of~\eqref{eq:ent_2pt}. 

First, the matrix elements~\eqref{eq:Gmat} are  obtained by using the 
Fourier transform~\eqref{eq:free_ft} as 
\begin{multline}
\label{eq:Gxy_expr1}
G_{x,y}(t)=\frac{1}{L}
\sum_{k,p}\langle\Phi_0|c_k^\dagger c_p 
|\Phi_0\rangle e^{i(kx-py)}e^{i(\varepsilon(k)-\varepsilon(p))t}=\\
\frac{1}{L}\sum_{k,j} \langle\Phi_0|c_k^\dagger c_{p_j(k)} |\Phi_0\rangle 
e^{i[kx-(k-\frac{\pi}{2}j)y]}e^{i[\varepsilon(k)-\varepsilon(k-\frac{\pi}{2}j)]t},
\end{multline}
where $k$ is the quasimomentum, $\varepsilon(k)$ is the dispersion of the 
$XX$ chain (cf.\eqref{eq:freechain_diag}), and $j=0,1,2,3$. 
In~\eqref{eq:Gxy_expr1} we exploited the $4$-site translation invariance 
of $|\Phi_0\rangle$, which implies that $\langle\Phi_0|c_k^\dagger c_p 
|\Phi_0\rangle\neq0$ only when $4(k-p)$ is an integer multiple of $2\pi$. 
Indeed, in~\eqref{eq:Gxy_expr1} we defined $p_j(k)$ as the 
quasimomentum  in $(-\pi,\pi]$ such that $k-p_j(k)=j\frac{\pi}{2} 
\mod 2\pi$. 

The expectation value in~\eqref{eq:Gxy_expr1} 
$\langle\Phi_0|c_k^\dagger c_{p_j(k)} |\Phi_0\rangle$ is given 
in~\eqref{eq:freeneg_G}. Thus in the thermodynamic limit $L\rightarrow\infty$, 
we can rewrite~\eqref{eq:Gxy_expr1} as 
\begin{multline}
\label{eq:Gxy_expr2}
G_{x,y}=\int_{-\pi}^\pi \frac{dk}{2\pi}e^{ik(x-y)}\Big[ \frac{1}{2}- i^y
\frac{1+i}{4}e^{it(\varepsilon(k)-\varepsilon(k-\frac{\pi}{2}))}\\
-(-i)^y\frac{1-i}{4} e^{it(\varepsilon(k)-\varepsilon(k-\frac{3\pi}{2}))}\Big].
\end{multline}
It is convenient to exploit explicitly the $4$-site periodicity 
of~\eqref{eq:Gxy_expr2}, defining the block matrix $\Gamma_{x,y}$ as 
\begin{equation}
\label{eq:Gxy_block}
\Gamma_{x,y}(t):=G_{4x+i,4y+j}=\\
\frac{1}{2}\int_{-\pi}^{\pi}\frac{dk}{2\pi} e^{4ik(x-y)}\Gamma_k, 
\end{equation}
where $\Gamma_k$ is defined as 
\begin{equation}
\Gamma_k:=\mathds{1}_4 + \Gamma^{(1)}_k e^{it(\varepsilon(k)-\varepsilon(k-\frac{\pi}{2}))} 
+ \Gamma^{(2)}_k e^{it(\varepsilon(k)-\varepsilon(k-\frac{3\pi}{2}))}, 
\end{equation}
and $\mathds{1}_4$ is the $4\times 4$ identity matrix, and we defined $\Gamma_k^{(1)}$ and 
$\Gamma_k^{(2)}$ as 
\begin{multline}
\label{eq:Gxy_block_expr}
		\Gamma^{(1)}_k
		=\frac{1}{2}\begin{pmatrix}
		1-i&(1+i)e^{-ik}\\
		(1-i)e^{ik}&1+i
		\end{pmatrix}
		\otimes
		\begin{pmatrix}
		1&-e^{-2ik}\\
		e^{2ik}&-1
		\end{pmatrix}=\\
		\frac{1}{2}(\mathbb{1}_{2}+\sigma^{(k)}_x-\sigma_y^{(k)}-i\sigma_z^{(k)})
		\otimes (\sigma_z^{(2k)}-i\sigma_y^{(2k)}). 
\end{multline}
We also defined 
\begin{multline}
\label{eq:Gxy_block_expr_1}
\Gamma^{(2)}_k 
=\frac{1}{2}\begin{pmatrix}
		1+i&(1-i)e^{-ik}\\
		(1+i)e^{ik}&1-i
		\end{pmatrix}
		\otimes
		\begin{pmatrix}
		1&-e^{-2ik}\\
		e^{2ik}&-1
		\end{pmatrix}=\\
		\frac{1}{2}(\mathds{1}_{2}+\sigma^{(k)}_x+\sigma_y^{(k)}+
		i\sigma_z^{(k)})\otimes (\sigma_z^{(2k)}-i\sigma_y^{(2k)}). 
\end{multline}
In~\eqref{eq:Gxy_block_expr} and~\eqref{eq:Gxy_block_expr_1} we introduced 
the rotated Pauli matrices $\sigma_\alpha^{(k)}$ as
	\begin{equation}
	\label{eq:pauli}
		\sigma_\alpha^{(k)}=e^{-i\frac{k}{2}\sigma_z} \; \sigma_\alpha \; e^{i\frac{k}{2}\sigma_z}.
	\end{equation}
Let us define the functions $f_1(k)$, $f_2(k)$ and $g(k)$ as 
\begin{align}
\label{eq:f1}
& f_1(k):=\mathbb{1}_{2}+\sigma^{(k)}_x-\sigma^{(k)}_y-i\sigma^{(k)}_z\\
\label{eq:f2}
& f_2(k):=\mathbb{1}_{2}+\sigma^{(k)}_x+\sigma_y^{(k)}+i\sigma^{(k)}_z\\
\label{eq:g}
& g(k):=\sigma^{(2k)}_z-i\sigma^{(2k)}_y.
\end{align}
Now, to evaluate $\mathrm{Tr}[G_A^n]$ one has to trace over the indices 
$x,y$ of the $\Gamma_{x,y}$  and also over products of the 
$4\times4$ blocks introduced in~\eqref{eq:Gxy_block}. 
The first trace can be performed by exploiting the identity
	\begin{equation}
	\label{eq:l_to_exp}
	\sum_{z=1}^{\ell/4} e^{4izk}=\frac{\ell}{8}\int_{-1}^{1} d\xi\; w([k]_{\pi/2}) e^{i(\ell \xi +\ell +4 )[k]_{\pi/2}/2}, 
	\end{equation}
where 
\begin{equation}
	w(k):=\frac{2k}{\sin(2k)}. 
\end{equation}
The notation $[k]_{\pi/2}$ in~\eqref{eq:l_to_exp} 
means that the quasimomentum $k$ is considered modulo $\pi/2$. 
Thus, we can rewrite $\mathrm{Tr}[G_A^n]$ as 
\begin{multline}
	\label{eq:Gn_1}
	\mathrm{Tr}[G_A^n]=
	\left(\frac{\ell}{8}\right)^n \int_{-\pi}^{\pi} \frac{d^n k}{(2\pi)^n}\int_{-1}^{1}d^n \xi \;
	\mathrm{Tr}\prod_{j=1}^{n}\Gamma_{k_j}\\
	\times\prod_{j=1}^{n} w([k_j-k_{j-1}]_{\pi/2})e^{i(\ell \xi_j +\ell +4 )[k_j-k_{j-1}]_{\pi/2}/2},
\end{multline}
where $\Gamma_k$ is the $4\times 4$ block matrix introduced  
in~\eqref{eq:Gxy_block}, and we  identified  $k_0\equiv k_n$. We are 
interested in finding the leading term in the hydrodynamic limit. The strategy is to 
use the stationary phase approximation of the integral  in~\eqref{eq:Gn_1}. 
The stationary phase approximation states that~\cite{wong} 
\begin{multline}
	\label{eq:st_phase}
\lim_{\ell\rightarrow\infty} \int_\Omega d^N x\; B(\mathbf{x}) e^{i\ell A(\mathbf{x})}=\\
\left(\frac{2\pi}{\ell}\right)^{N/2} \sum_j B(\mathbf{x_j}) |\det H(\mathbf{x_j})|^{-\frac{1}{2}}e^{i\ell A(\mathbf{x_j})+i\pi\sigma(\mathbf{x_j})/4},
\end{multline}
Here $A(\mathbf{x})$ and $B(\mathbf{x})$ are functions, and 
{$\mathbf{x_j}$} are the stationary points of $A(\mathbf{x})$ 
that are in the integraion domain $\Omega$. In~\eqref{eq:st_phase} $H$ is 
the Hessian matrix of $A$ and $\sigma$ is its signature, i.e.,  
the difference between the number of positive and negative eigenvalues. 
We now apply~\eqref{eq:st_phase} to the integral on the $2n-2$ variables 
$k_2, ...,k_n, \xi_2, ..., \xi_n$ in~\eqref{eq:Gn_1}. 
The stationarity conditions $\partial_{\xi_j} A=0$ 
imply that the stationary points must satisfy the equation 
\begin{equation}
	\label{eq:stat_xi}
	[k_j-k_1]_{\pi/2}=0 \qquad \qquad \forall j. 
\end{equation}
This implies that   $w([k_j-k_{j-1}]_{\pi/2})=1$ for 
all the stationary points. 

Let us now discuss the consequences of  the stationarity 
condition with respect to the $k_j$, i.e., $\partial_{k_j} A=0$. 
Now, the analysis is more complicated because one has to take the trace of 
arbitrary powers of $\Gamma_k$ (cf.~\eqref{eq:Gn_1}). Since 
$\Gamma_k$ is the sum of three terms, this means that for fixed $n$ there 
are $3^n$ terms. Moreover,  $\Gamma_k$ contains phase factors which have to be treated 
carefully in the stationary phase approximation. 

To proceed, we observe that both $\Gamma_k^{(1)}$ and $\Gamma_k^{(2)}$ contain 
a term $g(k)$ (cf.~\eqref{eq:g}). Moreover, due to the tensor product in~\eqref{eq:Gxy_block_expr} 
and~\eqref{eq:Gxy_block_expr_1}, we can perform the trace operation on the terms with $g(k)$ 
separately. In the following we discuss the conditions on $k_j$ to have a nonzero trace. 
We observe that:
\begin{itemize}
	\item[$(i)$] The terms $g(k)^2$ are identically zero. Since $g(k)=g(k\pm \pi)$, the terms 
	$g(k)g(k\pm\pi)$ are also zero. 
\item[$(ii)$] $g(k)g(k\pm \frac{\pi}{2})=2(\mathds{1}_{2}+\sigma^{(2k)}_x)$;
\item[$(iii)$] The trace of the product of an \textit{odd} number of 
	$g(k_i)$ with $[k_i-k_j]_{\pi/2}=0$ is zero. Indeed, one possibility is that 
	the product is identically zero, if two of the factors satisfy the condition in $(i)$.  
	The only other possibility is that the product is of the form 
	$a \sigma_y^{(2k)}+b\sigma_z^{(2k)}$, with $a,b$ constants. 
	This is obtained by repeatedly using $(ii)$. Again, the trace of the 
	result is zero. 
\item[$(iv)$] The trace of the product of an \textit{even} number $2m$ of blocks 
	$g(k_{j_i})$, provided that $k_{j_i}-k_{j_{i-1}}=\pm\pi/2\,\mathrm{mod}\,\,2\pi$, 
	is $2^{2m}$. This is a straightforward consequence of $(ii)$. 
\end{itemize}
In summary, the observations $(i)-(iv)$ imply that 
in the product $\prod_{j=1}^{n}\Gamma_{k_j}$ in~\eqref{eq:Gn_1} 
only terms with an even number $2m$ of factors $\Gamma^{(1)}_k$ or $\Gamma^{(2)}_k$ 
(cf.~\eqref{eq:Gxy_block_expr}  and~\eqref{eq:Gxy_block_expr_1}) are not zero.  
The quasimomenta $k_{j_1},...,k_{j_{2m}} \in (-\pi,\pi]$ associated to each $\Gamma^{(1)}$ 
or $\Gamma^{(2)}$ factor are such that $k_{j_i}-k_{j_{i-1}}=\pm\frac{\pi}{2} \mod 2\pi$. 
As discussed above, the trace over the factors $g(k)$ gives a factor $2^{2m}$. 
Let us now determine the contributions of the trace over the matrices  $f_1$ and 
$f_2$ (cf.~\eqref{eq:f1} and~\eqref{eq:f2}). To proceed, it is straightforward to 
check the following properties of $f_1$ and $f_2$ 
\begin{enumerate}
		\item $f_1(k)f_1(k+\frac{\pi}{2})=f_1(k)f_2(k+\frac{\pi}{2})=0$
		\item $f_2(k)f_1(k-\frac{\pi}{2})=f_2(k)f_2(k-\frac{\pi}{2})=0$
		\item $f_1(k)f_2(k-\frac{\pi}{2})=f_2(k)f_1(k+\frac{\pi}{2})=4(\mathbb{1}_{2}+\sigma^{(k)}_x)$ 
		\item $f_1(k)f_1(k-\frac{\pi}{2})=-4(\sigma_y^{(k)}+i\sigma_z^{(k)})$
		\item $f_2(k)f_2(k+\frac{\pi}{2})=4(\sigma_y^{(k)}+i\sigma_z^{(k)})$
		\item$f_1(k\pm\pi)f_2(k\pm\pi-\frac{\pi}{2})=f_2(k\pm\pi)f_1(k\pm\pi+\frac{\pi}{2})=
			4(\mathbb{1}_{2}-\sigma_x^{(k)})$
		\item $f_1(k\pm\pi)f_1(k\pm\pi-\frac{\pi}{2})=4(\sigma_y^{(k)}-i\sigma_z^{(k)})$
		\item $f_2(k\pm\pi)f_2(k\pm \pi+\frac{\pi}{2})=4(-\sigma_y^{(k)}+i\sigma_z^{(k)})$.
	\end{enumerate}
The first two relations show that, to have a nonzero product, if we have a matrix  
$f_1$ with an associated quasimomentum $k_{j_i}$, the next matrix must 
have an associated quasimomentum $k_{j_{i+1}}=k_{j_i}-\frac{\pi}{2}\; \text{mod}\; 2\pi$, 
while a  matrix $f_2$ with associated quasimomentum $k_{j_i}$ must be followed 
by a matrix with associated quasimomentum $k_{j_{i+1}}=k_{j_i}+\frac{\pi}{2}\; \text{mod}\; 2\pi$. 
Thus, $f_1$ can be seen as a ``lowering operator" for the 
quasimomentum and represented as $\searrow$. Similarly, $f_2$ can be seen as a 
``raising operator" and represented as $\nearrow$. 

We can represent any product of  $f_1$ and $f_2$ 
not yielding $0$ as a sequence of these operators, which raise or 
lower the starting $k_{i_1}=k$ by $\frac{\pi}{2}$. Again, we remind that  
we are interested only in \textit{even} sequences. 
Relations $3$ and $6$ are associated to the subsequences 
$\searrow\nearrow$ and $\nearrow\searrow$. 
Similarly, rules $4$ and $7$ are associated to the subsequence 
$\searrow\searrow$ and rules $5$ and $8$ to the subsequence 
$\nearrow\nearrow$. 

Moreover, from the rules $3$-$8$ it follows that only sequences 
with final quasimomentum equal to $k$ modulo $2\pi$ 
give a nonzero contribution, as an odd number of 
$\nearrow\nearrow$ or $\searrow\searrow$ (corresponding 
to a change of $\pm\pi$ in the quasimomentum) yield a 
product of the form $(a \sigma^{(k)}_y + b\sigma_z^{(k)})$, whose 
trace is zero. 

Finally, we have to compute the 
contribution of the  sequences of $f_1,f_2$  that give a nonzero result. 
To this purpose, let us observe that:
\begin{itemize}
	\item[$(a)$] a subsequence $\nearrow \searrow ... \nearrow \searrow$ 
		or $\searrow \nearrow... \searrow \nearrow$ yields 
		a factor $2^{3p/2-1} (\mathbb{1}_{2}+\sigma_x^{(k)})$, where 
		$p$ is the number of operators (rule $3$). 
		The same subsequence but with starting point $k\pm\pi$ 
		gives $2^{3p/2-1} (\mathbb{1}_{2}-\sigma_x^{(k)})$ 
		(see rule $6$). 
	\item[$(b)$] subsequences of four consecutive operators of the same 
		kind, i.e. $\nearrow\nearrow\nearrow\nearrow$ (rules $5$ and 
		$8$) or $\searrow\searrow\searrow\searrow$ (rules $4$ and $7$), 
		yield a factor $-2^5(\mathbb{1}_{2}\pm\sigma_x^{(k)})$. 
		The sign depends on whether the first operator is 
		associated with a quasimomentum $k$ (giving the $+$ sign) or 
		$k\pm\pi$ (giving the $-$ sign). 
	\item[$(c)$] any subsequence that can be decomposed in sub-blocks as those 
		described in $(a)$ and $(b)$ yields a factor 
		$(-1)^w \;2^{3p/2-1} (\mathbb{1}_{2}\pm\sigma_x^{(k)})$, 
		where $p$ is the number of operators, $w$ is the number of 
		``windings" around the Brillouin zone of the sequence, 
		and the sign depends on whether the first operator is 
		associated with a quasimomentum $k$ or $k\pm\pi$, as in $(b)$. 
	\item[$(d)$] A generic sequence cannot always be decomposed only in terms of subsequences of 
		the type in $(c)$. Let us consider a  \textquotedblleft maximal\textquotedblright subsequence of type $(c)$ that 
		can be identified in the main sequence (i.e., has not adjacent blocks of the form 
		$\nearrow\searrow$, $\searrow \nearrow$,$\nearrow\nearrow\nearrow\nearrow$ 
	or $\searrow\searrow\searrow\searrow$), and that starts from $k\pm\pi$. We represent 
	such sequence with a $\square$. It is clear that the $\square$ must be connected to 
	the remaining parts of the sequence as $\nearrow \nearrow \square \searrow \searrow$, 
	$\searrow \searrow \square \nearrow \nearrow$, 
	$\nearrow \nearrow \square \nearrow \nearrow$, or $\searrow \searrow \square \searrow \searrow$.
	This subsequence corresponds to a factor (see rules $4-8$) 
	$(-1)^w 2^{3p/2-1} (\mathbb{1}_{2}+\sigma_x^{(k)})$, where $p$ is 
	the number of operators, $w$ is the number of ``windings'' around the Brillouin 
	zone of the subsequence. Notice that in the cases $\nearrow\nearrow\square\nearrow\nearrow$ and  
	$\searrow\searrow\square\searrow\searrow$ the number of windings of $\square$ is raised by $1$. 
\item[$(e)$] By using rules $(a)-(d)$ we are left with sequences of the form 
	$\propto(\mathbb{1}_{2}+\sigma_x^{(k)})$.  Then, 
	it is straightforward  to realize that the contribution 
	of \textit{any} sequence that does not give zero 
	(that is, those with an integer number of ``windings" around the Brillouin zone) is 
	$(-1)^w \;2^{3p/2}$, where $p=2m$ is the (even) number of operators and $w$ the number of windings. 
	\end{itemize}
The result in $(e)$ allows us to write an expression that generates the 
contributions of all the sequences. Indeed, if we associate $f_1 \leftrightarrow 
\searrow\leftrightarrow2\sqrt{2}e^{-i\frac{\pi}{4}}$ and 
$f_2 \leftrightarrow \nearrow\leftrightarrow2\sqrt{2}e^{i\frac{\pi}{4}}$, we obtain 
that the  total contribution of the sequences with $p=2m$ factors is given by 
\begin{equation}
\label{eq:genfunc}
\mathrm{Tr}(f_1+f_2)^p=2^{\frac{3p}{2}}\mathrm{Re}\left[(e^{-i\frac{\pi}{4}}+ e^{i\frac{\pi}{4}})^p\right]. 
\end{equation}
In~\eqref{eq:genfunc} we used  that at any stationary point that gives a 
nonzero contribution the oscillating factors that are present in the rotated 
Pauli matrices (cf.~\eqref{eq:pauli})  cancel out. 

Let us now proceed to determine the consequences of the stationarity  conditions 
with respect to $k_j$. The generic term originating from the product 
$\prod_{j=1}^{n}\Gamma_{k_j}$ (cf.~\eqref{eq:Gxy_block}) contains $f_1(k_j)$, $f_{2}(k_j)$, or 
the identity. In the last case the quasimomentum $k_j$ does not appear. Stationarity with 
respect to the missing quasimomenta $k_j$ imply that 
\begin{equation}
\label{eq:stat_k_id}
		\ell(\xi_{j}-\xi_{j+1})=0 \quad \Rightarrow \quad \xi_{j}=\xi_{j+1}.
\end{equation}
Instead, for the quasimomenta $k_{j_1}, ..., k_{j_m}$ that 
appear in the block the stationarity condition yields 
\begin{multline}
\label{eq:stat_k_exps}
\ell (\xi_{j_i} - \xi_{j_{i}+1}) + t \Big[v(k_{j_i})-v\Big(k_{j_i}\pm\frac{\pi}{2}\Big)\Big]=0 \\ 
\Rightarrow \quad \ell \xi_{j_i} + v(k_{j_i}) t =\ell \xi_{j_{i+1}} + v(k_{j_{i+1}}) t \quad \forall i. 
\end{multline}
In~\eqref{eq:stat_k_exps} we used that  $\xi_{j_{i}+1}= \xi_{j_{i+1}}$, which  
follows from~\eqref{eq:stat_k_id} applied to the $\xi_l$ with $j_i<l<j_{i+1}$.  
Moreover, the condition to have a nonzero trace implies that 
$k_{j_i}\pm\frac{\pi}{2}=k_{j_{i+1}}$. 
The conditions~\eqref{eq:stat_k_id}~\eqref{eq:stat_k_exps} give some 
non-trivial constraints on the stationary value  of $\xi_1=\xi_{j_m}$. 
The constraint is determined by the condition that 
all the $\xi_j\in [-1,1]$.  Let us define $k:=k_{j_n}$. The result 
depends on the remaining 
quasimomenta in the string of operators $\Gamma_{k_j}$. 

We can distinguish three families of 
quasimomenta $k_{j_i}$ 
\begin{enumerate}
	\item[$(\alpha)$]\label{case-alpha} The quasimomenta in the string take only the values $k,k+\frac{\pi}{2}$ or $k,k-\frac{\pi}{2}$. 
	\item[$(\beta)$]\label{case-beta} The quasimomenta $k_{j_i}$ take only three of the four values $k$, 
	$k+\frac{\pi}{2}$, $k-\frac{\pi}{2}$, $k\pm\pi$. 
\item[$(\gamma)$]\label{case-gamma} The quasimomenta $k_{j_i}$ take all the four values $k$, $k+\frac{\pi}{2}$, $k-\frac{\pi}{2}$, $k\pm\pi$.
\end{enumerate}
Now, one can verify that for case $(\alpha)$, Eq.~\eqref{eq:stat_k_exps} 
implies the condition
\begin{equation}
\label{eq:xi-step}
\xi_{j_i}= \xi_1 + \Big[v(k)-v(k\pm\frac{\pi}{2})\Big]\frac{t}{\ell} \in [-1,1]. 
\end{equation}
Thus, the integration over $\xi_1$ gives the function $M_1(k)$ 
\begin{equation}
\label{eq:cin_case1}
M_1(k)=\max\Big\{0,2-\Big|v(k)-v\Big(k\pm\frac{\pi}{2}\Big)\Big|
\frac{t}{\ell}\Big\},
\end{equation}
where $v(k)=2\sin(k)$. 
For  the second case $(\beta)$, without loss of generality we can choose $k$ to 
be the smaller of the three quasimomenta present. 
The constraints from~\eqref{eq:stat_k_exps} now read 
\begin{align}
	\label{eq:cin_case2_1}
		& \xi_{j_i}= \xi_1 + \Big[v(k)-v\Big(k+\frac{\pi}{2}\Big)\Big]\frac{t}{\ell} \in [-1,1],\\
		& \xi_{j_j}= \xi_1 + \Big[v(k)-v(k+\pi)\Big]\frac{t}{\ell} \in [-1,1].
\end{align}
By integrating over $\xi_1$, a straightforward but tedious  calculation yields 
\begin{multline}
\label{eq:cin_case2_2}
M_2(k)=\max \Big\{0,2-\max\Big\{\Big|v(k)-v\Big(k+\frac{\pi}{2}\Big)\Big|,\\
|v(k)-v(k+\pi)|,\Big|v(k)+v\Big(k+\frac{\pi}{2}\Big)\Big|\Big\}\frac{t}{\ell}\Big\}.
\end{multline}
Finally, in the third case $(\gamma)$ we have three constraints like those in 
equations~\eqref{eq:cin_case1},~\eqref{eq:cin_case2_1}. Specifically,  we obtain 
\begin{multline}
\label{eq:cin_case3}
M_3(k)=\max \{0,2-\max\{|v(k)-v(k+\pi)|,\\
\Big|v\Big(k-\frac{\pi}{2}\Big)+v\Big(k+\frac{\pi}{2}\Big)\Big|\Big\}\frac{t}{\ell}\Big\}.
\end{multline}
Before putting everything together, we notice that for the stationary points that 
give a nonzero contribution we have $|\det H(\mathbf{x_j})|=(\frac{1}{2})^{2n-2}$ 
and $\sigma(\mathbf{x_j})=0$ (cf.~\eqref{eq:st_phase}). Finally, we obtain 
\begin{widetext}
\begin{multline}
	\label{eq:Gn_final_1}
	\mathrm{Tr}[G_A^n]=\frac{\ell}{16^n}\int_{-\pi}^{\pi}\frac{dk}{2\pi}\; 2^{n-1} \Big\{2\cdot 4^n 
	+ \sum_{m=1}^{\lfloor\frac{n}{2}\rfloor} \binom{n}{2m} 4^{n-2m} 
	\Big(\frac{1}{2}\Big)^{2m}2^{2m}\cdot 2^{3m}\Big[ 2 M_1(k) + 
		4 \sum_{j=1}^{\lfloor\frac{m}{2}\rfloor}\binom{m}{2j} M_2(k)+\\
	+\Big((e^{i\frac{\pi}{4}}+e^{-i\frac{\pi}{4}})^{2m}-2-
	4\sum_{j=1}^{\lfloor\frac{m}{2}\rfloor}\binom{m}{2j}\Big) M_3(k)\Big]\Big\},
\end{multline}
\end{widetext}
where $M_1, M_2, M_3$ are the kinematic terms~\eqref{eq:cin_case1},~\eqref{eq:cin_case2_2},~\eqref{eq:cin_case3}. 
Let us explain the various combinatorial factors in~\eqref{eq:Gn_final_1}. The powers of $4$ account for the 
four choices $k\pm\pi/2,k\pm\pi$ that one has for the quasimomenta that are missing in the string 
$\Gamma_{k_{j_1}}\Gamma_{k_{j_2}}\cdots\Gamma_{k_{j_n}}$. 
The missing quasimomenta are obtained by selecting  $\mathbb{1}_{4}$ in~\eqref{eq:Gxy_block}. For instance, 
the term $2\cdot 4^n$ in~\eqref{eq:Gn_final_1} is the contribution in which all the $\Gamma_{k_j}$ are replaced 
by $\mathds{1}_4$. Notice that the factor $2$ in $2\cdot 4^{n}$ comes from the integral over $\xi_1$. 
The binomial $\binom{n}{2m}$ in the second term in~\eqref{eq:Gn_final_1} counts the possible ways 
to choose an even subsequence of $\Gamma_{k_{j_1}}\Gamma_{k_{j_2}}\cdots\Gamma_{k_{j_{2m}}}$.  Each of them 
gives a factor $\frac{1}{2}$ (cf.~\eqref{eq:Gxy_block}).  Moreover, there is a factor $2^{2m}$  and 
$2^{3m}$ from the trace in rule $(iv)$ and from~\eqref{eq:genfunc}, respectively. Let us now discuss the 
term within the square brackets in~\eqref{eq:Gn_final_1}. The three terms in the square brackets corresponds 
to the three cases $(\alpha,\beta,\gamma)$. The first term corresponds to case $(\alpha)$, in which the quasimomenta 
in the string can have only the values $k\pm\pi/2$. Now, there are only the two cases 
($\nearrow\searrow ... \nearrow\searrow$ and $\searrow\nearrow ...\searrow\nearrow$) to consider, 
each of them giving $1$, and the factor $M_1(k)$. The second term in the square brackets corresponds to 
case $(\beta)$, in which we have configurations with an even number $2j$ of (alternated) pairs $\searrow\searrow$ 
and $\nearrow\nearrow$, univocally connected by subsequences of the type (a). Each configuration contributes with 1, and we have  
$4 \sum_{j=1}^{\lfloor\frac{m}{2}\rfloor}\binom{m}{2j}$ of such configurations. The summation accounts for the ways 
where to place the pairs $\searrow\searrow$ and $\nearrow\nearrow$ after the string of $2m$ operators
has been divided in $m$ slots of two.
Moreover, the partition of the string of operators can be done by starting from even or odd sites of the string, which gives a factor $2$. 
Besides that, there is another factor $2$, coming from the fact that one can put either $\searrow\searrow$ 
or $\nearrow\nearrow$ in the first chosen slot, the others being filled accordingly. 
Finally, the remaining contributions are obtained by subtracting the cases  
described above from the total $(e^{i\frac{\pi}{4}}+e^{-i\frac{\pi}{4}})^{2m}$ (see~\eqref{eq:genfunc}). 
It is now straightforward to simplify equation~\eqref{eq:Gn_final_1} to obtain 
\begin{multline}
\label{eq:Gn_final_2}
\mathrm{Tr}[G_A^n]=\frac{\ell}{2} \\
+ 
\ell \int_{-\pi}^{\pi}\frac{dk}{2\pi}
\Big\{\Big[2\Big(\frac{1}{2}\Big)^n-\Big(\frac{2+\sqrt{2}}{4}\Big)^n-\Big(\frac{2-\sqrt{2}}{4}\Big)^n\Big]
m_1(k,t)+\\
+\Big[2\Big(\frac{2+\sqrt{2}}{4}\Big)^n+2\Big(\frac{2-\sqrt{2}}{4}\Big)^n-2\Big(\frac{1}{2}\Big)^n-
1\Big]m_2(k,t)+\\
+\Big[\frac{1}{2}+\Big(\frac{1}{2}\Big)^n-\Big(\frac{2+\sqrt{2}}{4}
\Big)^n-\Big(\frac{2-\sqrt{2}}{4}\Big)^n\Big]m_3(k,t)\Big\},
\end{multline}
where we have defined:
\begin{equation}
	\label{eq:mins}
	m_1(k,t)=\min\left\{1,\frac{t}{\ell} |v(k)-v(k+\frac{\pi}{2})|\right\},
\end{equation}
\begin{multline}
	m_2(k,t)=\min\Big\{1,\frac{t}{\ell}\max\Big\{|v(k)-v(k+\frac{\pi}{2})|,\\
	|v(k)-v(k+\pi)|,|v(k)-v(k-\frac{\pi}{2})|\Big\}\Big\},
\end{multline}
\begin{multline}
	 m_3(k,t)=\min \Big\{1,\frac{t}{\ell}\max\Big\{|v(k)-v(k+\pi)|,\\
	 |v(k-\frac{\pi}{2})-v(k+\frac{\pi}{2})|\Big\}\Big\}.
\end{multline}
Having the hydrodynamic prediction for $\mathrm{Tr}[G_A^n]$  for any $n$ allows 
us to obtain the prediction for the von Neumann entropy $S_A$. The strategy is 
to write $S_A=\mathrm{Tr}f(G_A)$, with $f(x)$ as defined in~\eqref{eq:quadr_abc}. 
After expanding $f(x)$ around $x=0$, and using~\eqref{eq:Gn_final_2} together with  
$f(0)=f(1)=0$ and $f(x)=f(1-x)$), we obtain 
\begin{multline}
\label{eq:f(G)}
S_{A}(t)
=\ell \int_{-\pi}^{\pi}\frac{dk}{2\pi} 
\Big[\Big(2f\Big(\frac{1}{2}\Big)-2f\Big(\frac{2+\sqrt{2}}{4}\Big)\Big)m_1(k,t)+\\
+\Big(4f\Big(\frac{2+\sqrt{2}}{4}\Big)-2f\Big(\frac{1}{2}\Big)\Big)
m_2(k,t)\\
+\Big(f\Big(\frac{1}{2}\Big)-2f\Big(\frac{2+\sqrt{2}}{4}\Big)\Big)m_3(k,t)\Big].
\end{multline}
Let us now show that the \emph{ab initio} result~\eqref{eq:f(G)} coincides with 
the result obtained from the method introduced in section \ref{sec:genmethod}. 
The latter approach yields 
\begin{widetext}
\begin{multline}
\label{eq:mostro}
S_A(t)=\int_{\pi/2}^{3\pi/4} \frac{dk}{2\pi}\Big\{(s_{\{1\}}+s_{\{3\}})
\Big[(v_1-v_2)t\; \Theta(\ell-(v_1-v_2)t) + \ell \Theta((v_1-v_2)t-\ell)+(v_4-v_3)t\;\Theta(\ell-(v_1-v_3)t)+\\
+(\ell-(v_1-v_4)t)\chi\big(\ell/({v_1 t-v_3 t}),\ell/(v_1 t-v_4 t)\big)\Big]\\
+(s_{\{2\}}+s_{\{4\}})\Big[ ((v_1-v_4)t-\ell)\chi\big(\ell/(v_1 t-v_4 t),\min\{\ell/(v_1 t-v_2 t),\ell/(v_2 t-v_4 t)\}\big)+
	(v_1-v_2) t \chi\big(\ell/(v_2 t-v_4 t),\ell/(v_1 t-v_2 t)\big)\\
+(v_2-v_4)t\chi\big(\ell/(v_1 t-v_2 t),\ell/(v_2 t-v_4 t)\big)+\ell 
\Theta\big(t-\max\big\{\ell/(v_1 t-v_2 t),\ell/(v_2 t-v_4 t)\big\}\big)\Big]
	+\\+(s_{\{1,2\}}+s_{\{3,4\}})\Big[(v_2-v_4)t\Theta(\ell-(v_1-v_4)t)+(\ell-(v_1-v_2)t)+
	\chi\big(\ell/(v_1 t-v_4 t),\ell/(v_1 t-v_2 t)\big)\Big]+\\
+s_{\{1,3\}}\Big[((v_1-v_3)t-\ell)\chi\big(\ell/(v_1 t-v_3 t),\ell/(v_1 t-v_3 t)\big)+(\ell-(v_2-v_4)t)\chi\big(\ell/(v_1 t-v_4 t),
\ell/(v_2 t-v_4 t)\big)\Big]\\
+\int_{3\pi/4}^{\pi}\frac{dk}{2\pi} \{1\leftrightarrow 2,3 \leftrightarrow 4\}. 
\end{multline}
\end{widetext}
Here $\Theta(x)$ is the Heaviside theta function, and 
$\chi(a,b)$ is the characteristic function of the interval $[a,b]$, 
with the caveat that if $b<a$, it is zero. In~\eqref{eq:mostro} we 
dropped the dependence of the velocities on $k$ for the sake of clarity. 
The first term in~\eqref{eq:mostro} corresponds to the situation with quasiparticle 
$1$ or $3$ in subsystem $A$ or $\overline{A}$. The second term describes the case with quasiparticles 
$2$ or $4$ in $A$ or $\overline{A}$. The third and fourth terms take into account the situations with two 
quasiparticles in $A$ and two in $\overline{A}$. The last term in~\eqref{eq:mostro} is obtained from the 
previous ones by exchanging $1\leftrightarrow 2$ and $3\leftrightarrow 4$. 
To proceed, we determine the contributions $s_{\{x\}}$ in~\eqref{eq:mostro}. 
These are obtained from~\eqref{eq:freeneg_G} by using the strategy described in 
section~\ref{sec:genmethod}. A straightforward calculation gives 
$s_{\{1\}}=s_{\{2\}}=s_{\{3\}}=s_{\{4\}}=f(1/2), s_{\{1,2\}}=s_{\{3,4\}}=
2f((2+\sqrt{2})/4), s_{\{1,3\}}=2f(1/2)$. Now, it is straightforward, although tedious, to 
check that~\eqref{eq:mostro} is exactly the same as~\eqref{eq:f(G)}.

\bibliography{bibliography}	
\end{document}